\documentclass[aps,floats]{revtex4-2}
\usepackage{amsmath,amssymb}
\usepackage{graphicx,epsfig}
\usepackage[greek,english]{babel}
\usepackage{bbold}

\begin{document}
\bibliographystyle {plain}

\pdfoutput=1
\def\oppropto{\mathop{\propto}} 
\def\opsimeq{\mathop{\simeq}}
\def\opoverderline{\mathop{\overline}}
\def\operarrow{\mathop{\longrightarrow}}
\def\opsim{\mathop{\sim}}

\def\opmin{\mathop{\min}} 
\def\opmax{\mathop{\max}} 
\def\oplim{\mathop{\lim}}

\title{ Inverse problem in the conditioning of Markov processes on trajectory observables :  \\
what canonical conditionings can connect two given Markov generators ?  } 


\author{C\'ecile Monthus}
\affiliation{Universit\'e Paris-Saclay, CNRS, CEA, Institut de Physique Th\'eorique, 91191 Gif-sur-Yvette, France}


\begin{abstract}
In the field of large deviations for stochastic dynamics, the canonical conditioning of a given Markov process with respect to a given time-local trajectory observable over a large time-window has attracted a lot of interest recently. In the present paper, we analyze the following inverse problem: when two Markov generators are given, is it possible to connect them via some canonical conditioning and to construct the corresponding time-local trajectory observable? We focus on continuous-time Markov processes and obtain the following necessary and sufficient conditions: (i) for continuous-time Markov jump processes, the two generators should involve the same possible elementary jumps in configuration space, i.e. only the values of the corresponding rates can differ; (ii) for diffusion processes, the two Fokker-Planck generators should involve the same diffusion coefficients, i.e. only the two forces can differ. In both settings, we then construct explicitly the various time-local trajectory observables that can be used to connect the two given generators via canonical conditioning. This general framework is illustrated with various applications involving a single particle or many-body spin models. In particular, we describe several examples to show how non-equilibrium Markov processes with non-vanishing steady currents can be interpreted as the canonical conditionings of detailed-balance processes with respect to explicit time-local trajectory observables.

\end{abstract}

\maketitle


\section{ Introduction }

The theory of large deviations has become the unifying language for statistical physics
 (see the reviews \cite{oono,ellis,review_touchette} and references therein)
 and has played a major role in the recent achievements obtained in the field of nonequilibrium 
(see the reviews with different scopes \cite{derrida-lecture,harris_Schu,searles,harris,mft,sollich_review,lazarescu_companion,lazarescu_generic,garrahan_lecture,jack_review}, 
the PhD Theses \cite{fortelle_thesis,vivien_thesis,chetrite_thesis,wynants_thesis,chabane_thesis,duBuisson_thesis} 
 and the Habilitation Thesis \cite{chetrite_HDR}).
In particular, the large deviations properties of time-local trajectory observables over a large time-window
have attracted a lot of interest in many different physical contexts 
 \cite{peliti,derrida-lecture,sollich_review,lazarescu_companion,lazarescu_generic,jack_review,vivien_thesis,chabane_thesis,duBuisson_thesis,lecomte_chaotic,lecomte_thermo,lecomte_formalism,lecomte_glass,kristina1,kristina2,jack_ensemble,simon1,simon2,tailleur,simon3,Gunter1,Gunter2,Gunter3,Gunter4,chetrite_canonical,chetrite_conditioned,chetrite_optimal,chetrite_HDR,touchette_circle,touchette_langevin,touchette_occ,touchette_occupation,garrahan_lecture,Vivo,c_ring,c_detailed,chemical,derrida-conditioned,derrida-ring,bertin-conditioned,touchette-reflected,touchette-reflectedbis,c_lyapunov,previousquantum2.5doob,quantum2.5doob,quantum2.5dooblong,c_ruelle,lapolla,c_east,chabane,us_gyrator,duBuisson_gyrator,duBuisson_area,c_largedevpearson}
 with the construction of the corresponding canonical conditioned processes (see the two very detailed papers \cite{chetrite_conditioned,chetrite_optimal} and references therein).
 However in most physical applications, their scaled-cumulant generating functions
 and the corresponding canonical conditioned generators
 are unfortunately not explicit, since one needs to be able to solve eigenvalue equations
 for appropriate deformations of the initial generators.
 It is thus useful to consider the large deviations at the so-called level 2.5 
 concerning the joint distribution of the empirical density and of the empirical flows,
  where it is possible to write explicit rate functions for arbitrary Markov generators.
The most important Markov processes where these explicit rate functions at level 2.5 have been much studied
are the discrete-time Markov chains 
\cite{fortelle_thesis,fortelle_chain,review_touchette,c_largedevdisorder,c_reset,c_inference,c_microcanoEnsembles,c_diffReg},
the continuous-time Markov jump processes with discrete space
\cite{fortelle_thesis,fortelle_jump,maes_canonical,maes_onandbeyond,wynants_thesis,chetrite_formal,BFG1,BFG2,chetrite_HDR,c_ring,c_interactions,c_open,barato_periodic,chetrite_periodic,c_reset,c_inference,c_LargeDevAbsorbing,c_microcanoEnsembles,c_susyboundarydriven,c_diffReg},
the diffusion processes in continuous space
\cite{wynants_thesis,maes_diffusion,chetrite_formal,engel,chetrite_HDR,c_lyapunov,c_inference,c_susyboundarydriven,c_diffReg}, 
as well as jump-diffusion or jump-drift processes \cite{c_reset,c_runandtumble,c_jumpdrift,c_SkewDB}.
Since any time-local trajectory observable can be rewritten in terms of the empirical density and of the empirical flows
with appropriate coefficients, its rate function can be obtained from
the optimization of the explicit rate function at level 2.5 over the empirical density and of the empirical flows
with appropriate constraints, but the solution of this optimization problem is unfortunately not explicit either in most applications.

Since the direct problem of the canonical conditioning of a given Markov process with respect to a given time-local trajectory observable is usually not explicit, the goal of the present paper is to gain some insight via
the analysis of the following inverse problem: when two Markov generators are given, is it possible to connect them via some canonical conditioning and to construct the corresponding time-local trajectory observable?
We will focus on continuous-time Markov processes, either Markov jump processes in discrete configuration space
or diffusions processes in continuous space and obtain explicit solutions for this inverse problem. 
This general framework will be illustrated with specific examples involving either a single particle or many-body spin models. 
In particular, since the issue of whether non-equilibrium dynamics can be considered as some conditionings of equilibrium dynamics has been much debated recently \cite{Evans2004,Evans2005,Evans2008,Baule2008,Evans2010,JackSollich2010,c_maximizationDynEntropy,Andrieux2012,JackEvans,chetrite_optimal,verley2016,MaximumCaliber,verley2022,Andrieux2022},
we describe various applications where non-equilibrium Markov processes with non-vanishing steady currents can be interpreted as the canonical conditioning of detailed-balance dynamics with respect to explicit time-local trajectory observables.

For clarity, the present paper is divided into two separate parts :

$\bullet$ The main text is devoted to Markov jump processes in discrete configuration space.
In Section \ref{sec_jump}, we recall the direct problem of the canonical condition of a given Markov jump process
with respect to time-local trajectory observables in relation with the dynamical large deviations 
of the empirical density and of the empirical flows.
In section \ref{sec_Inverse}, we consider the inverse problem where two Markov generators 
with the same possible transitions are given
and we construct explicitly the various canonical conditionings that can connect them.
To illustrate this general framework, we first consider  
a single particle on a one-dimensional periodic ring with directed dynamics (Section \ref{sec_Directed}) or undirected dynamics (Section \ref{sec_UnDirected}),
and we then turn to many-body spin models with single-spin dynamics (Section \ref{sec_SingleSpinFlip}) or two-spin-flip dynamics (Section \ref{sec_TwoSpinFlip}).
Our conclusions are summarized in section \ref{sec_conclusion}.

$\bullet$ The appendices are devoted to diffusion processes in dimension $d$.
In Appendix \ref{app_Diffusion}, we recall the direct problem of the canonical condition of a given diffusion process
with respect to time-local trajectory observables in relation with the dynamical large deviations of the empirical density and the empirical current.
In Appendix \ref{app_Inverse}, we consider the inverse problem where two Fokker-Planck generators with the same diffusion coefficients are given
and we construct explicitly the various canonical conditionings that can connect them.
This general framework is then illustrated with examples on the one-dimensional periodic ring (Appendix \ref{app_Ring})
and in the full space $R^d$ in dimension $d$ (Appendix \ref{app_Diffd}).


\section{ Reminder on the canonical conditioning for Markov jump processes } 

\label{sec_jump}

In this Section, we recall the direct problem of the canonical condition of a given Markov jump process with generator $w$
with respect to trajectory observables in relation with the dynamical large deviations of the empirical density and the empirical flows.

\subsection{ Properties of the Markov generator $w(.,.)$ with its steady state $P_*(.)$ } 

The dynamics for the probability $P_t(x) $ to be in configuration $x$ at time $t$
\begin{eqnarray}
\partial_t P_t(x) =    \sum_{y }   w(x,y)  P_t(y) 
\label{mastereq}
\end{eqnarray}
is governed by the Markov matrix $w(x,y) $ :
the off-diagonal $x \ne y$ positive matrix element $w(x,y) \geq 0 $  represents the jump rate 
per unit time from configuration $y$ towards the other configuration $x \ne y$,
while the negative diagonal elements can be computed in terms of the off-diagonal elements  
\begin{eqnarray}
w(y,y)   =  - \sum_{x \ne y} w(x,y) 
\label{wdiag}
\end{eqnarray}
in order to conserve the normalization of the total probability 
\begin{eqnarray}
 \sum_x P_t(x)  =  1
\label{markovchaincnormaconserved}
\end{eqnarray}
In this paper we will always assume that the dynamics of Eq. \ref{mastereq}
converges towards some normalizable steady-state $P_*(y)$ 
satisfying Eq. \ref{mastereq}
\begin{eqnarray}
0 =     \sum_{y }   w(x,y)  P_*(y) 
\label{mastereqst}
\end{eqnarray}

From the point of view of the spectral properties of the Markov matrix $w(.,.)$,
Eqs \ref{wdiag} and \ref{mastereqst} mean that 
 the highest eigenvalue of the Markov matrix $w(.,.)$ is vanishing 
 \begin{eqnarray}
  0  && = \langle l \vert w = \sum_x l(x) w(x,y)
 \nonumber \\
  0   && =  w \vert r \rangle =  \sum_{y }   w(x,y)  r(y) 
\label{mastereigen0}
\end{eqnarray}
where the positive left eigenvector associated to the conservation of probability is trivial 
\begin{eqnarray}
 l(x)=1
\label{markovleft}
\end{eqnarray}
while the positive right eigenvector is given by the steady state
\begin{eqnarray}
 r(y)=P_*(y) 
\label{markovright}
\end{eqnarray}


\subsection{ Spontaneous fluctuations of time-averaged observables over a large-time window $[0,T]$  } 

Let us now recall how time-averaged observables over a large-time-window $[0,T]$ 
can fluctuate around their steady values.

\subsubsection{ Empirical probability $P(.)$ and empirical flows $Q(.,.)$ with their constitutive constraints} 

For a given long trajectory $x(0 \leq t \leq T) $ of the Markov jump process, it is useful to introduce :

(a) the empirical probability
\begin{eqnarray}
  P(x)  \equiv \frac{1}{T} \int_0^T dt \  \delta_{x(t),x}  
 \label{rho1pj}
\end{eqnarray}
that measures the fraction of the time spent by the trajectory $x(0 \leq t \leq T) $ in each 
discrete configuration $x$ 
with the normalization to unity
\begin{eqnarray}
\sum_x  P(x)   = 1
\label{rho1ptnormaj}
\end{eqnarray}

(b) the empirical flows 
\begin{eqnarray}
Q(x,y) \equiv  \frac{1}{T} \sum_{t \in [0,T] : x(t^+) \ne x(t^-)} \delta_{x(t^+),x} \delta_{x(t^-),y} 
\label{jumpempiricaldensity}
\end{eqnarray}
that represent the density of jumps
from one configuration $y$ towards the other configuration $x \ne y$ 
seen during the trajectory $x(0 \leq t \leq T) $.
For large $T$, these empirical flows satisfy the following stationarity constraints : 
for any configuration $x$, the total incoming flow into $x$ is equal to the total outgoing flow out of $x$
(up to boundary terms of order $1/T$ that involve only the initial configuration at $t=0$ and the final configuration at time $t=T$)
\begin{eqnarray}
\sum_{y \ne x} Q(x,y)= \sum_{y \ne x} Q(y,x) \ \ \ \text{ for any $x$}
\label{contrainteq}
\end{eqnarray}

The empirical probability $P(.) $ of Eq. \ref{rho1pj}
 and of the empirical flows $ Q(.,.) $ of Eq. \ref{jumpempiricaldensity}
are the basic building blocks that are useful to reconstruct any time-local trajectory observable
via its parametrization by two functions $\Omega(x)$ and $\Lambda(x , y)$
\begin{eqnarray}
{\cal O}^{[\Omega(.) ; \Lambda(.,.)]}\left[ x(0 \leq t \leq T) \right]  
&& =   \frac{1}{T}  \int_0^T dt  \Omega( x(t)) +  
  \frac{1}{T} \sum_{t \in [0,T] : x(t^+) \ne x(t^-) } \Lambda(x(t^+) , x(t^-)  ) 
  \nonumber \\
  &&   =  \sum_{x  } P(x)  \Omega( x)  +       \sum_{y  }\sum_{ x \ne y  }Q(x,y)   \Lambda(x , y)  
\label{observableQ}
\end{eqnarray}

In particular,
the probability ${\cal P }^{Traj}_{[w(.,.)]}\left[ x(0 \leq t \leq T) \right]  $ of the trajectory $x(0 \leq t \leq T) $
for the dynamics governed by the Markov generator $w(.,.)$
can be rewritten in terms of the empirical probability $P(.) $ 
 and of the empirical flows $ Q(.,.) $ of the trajectory
\begin{eqnarray}
{\cal P }^{Traj}_{[w(.,.)]}\left[ x(0 \leq t \leq T) \right]  
&& \equiv e^{ \displaystyle  \int_0^T dt  w(x(t) , x(t) )  
 + \sum_{t \in [0,T] : x(t^+) \ne x(t^-) } \ln ( w(x(t^+) , x(t^-) ) ) }
 \nonumber \\
 && = e^{ \displaystyle T \left[ \sum_x P(x)w(x,x)  + \sum_y \sum_{x \ne y}  
  Q(x,y) \ln ( w(x,y ) ) \right] }
\label{pwtrajjump}
\end{eqnarray}


\subsubsection{ Explicit joint distribution of the empirical probability $P(.)$ and the empirical flows $Q(.,.)$ for large $T$} 

The joint probability distribution $P^{[2.5]}_T \left[ P(.)  ; Q(.,.) \right] $ of the empirical observables $\left[ P(.)  ; Q(.,.) \right] $ follows the large deviation form for large $T$
\cite{fortelle_thesis,fortelle_jump,maes_canonical,maes_onandbeyond,wynants_thesis,chetrite_formal,BFG1,BFG2,chetrite_HDR,c_ring,c_interactions,c_open,barato_periodic,chetrite_periodic,c_reset,c_inference,c_LargeDevAbsorbing,c_microcanoEnsembles,c_susyboundarydriven,c_diffReg}
\begin{eqnarray}
P^{[2.5]}_T \left[ P(.)  ; Q(.,.) \right] \opsimeq_{T \to +\infty} 
C_{2.5}\left[ P(.)  ; Q(.,.) \right] e^{- T I_{2.5}\left[ P(.)  ; Q(.,.) \right] }
\label{level2.5master}
\end{eqnarray}
where $C_{2.5}\left[ P(.)  ; Q(.,.) \right]$
summarizes the constitutive constraints 
Eqs \ref{rho1ptnormaj} and \ref{contrainteq}
\begin{eqnarray}
C_{2.5}\left[ P(.)  ; Q(.,.) \right]
  = \delta \left( \sum_x  P(x) - 1 \right) 
 \prod_x \delta \left( \sum_{y \ne x} [ Q(x,y)-  Q(y,x) ] \right)
\label{constraints2.5master}
\end{eqnarray}
while the rate function at Level 2.5 reads
\begin{eqnarray}
I_{2.5}\left[ P(.)  ; Q(.,.) \right]
&&  = \sum_{y } \sum_{x \ne y} 
\left[ Q(x,y)  \ln \left( \frac{ Q(x,y)  }{  w(x,y) P(y)  }  \right) 
 - Q(x,y)  + w(x,y)  P(y)  \right]
\label{rate2.5master}
\end{eqnarray}


\subsubsection{ Different perspective when the empirical flows $Q(.,.)$ 
are replaced by the empirical Markov generator $w^E(.,.)$  }

A useful different perspective can be obtained via the replacement of the empirical flows $Q(.,.)$ by
the empirical Markov generator $w^E(.,.)$ with the off-diagonal elements
\begin{eqnarray}
{\rm for } \ \ x \ne y : \ \ w^E(x,y)   =  \frac{ Q(x,y) }{P(y)} 
= \frac{\displaystyle  \sum_{t \in [0,T] : x(t^+) \ne x(t^-)} \delta_{x(t^+),x} \delta_{x(t^-),y} }{ \int_0^T dt \  \delta_{x(t),y}  }
\label{woffdiagempi}
\end{eqnarray}
while the diagonal elements are determined by the conservation of probability of Eq. \ref{wdiag}
\begin{eqnarray}
w^E(y,y)   =  - \sum_{x \ne y} w^E(x,y) 
\label{wdiagempi}
\end{eqnarray}
Then the constitutive constraint of Eq. \ref{contrainteq} for the flows $Q(.,.)$ translates into
\begin{eqnarray}
0 = \sum_{y \ne x} w^E(x,y) P(y)- P(x) \sum_{y \ne x} w^E(y,x) 
=  \sum_{y \ne x} w^E(x,y) P(y)+ w^E(y,y) P(x) =  \sum_y w^E(x,y) P(y)
\label{contrainteqinfer}
\end{eqnarray}
i.e. the empirical density should be the steady state of the empirical generator $w^E(.,.)$.

Then the joint distribution of Eq. \ref{level2.5master}
translates into the following joint distribution of the empirical probability $P(.)$ and of the empirical generator $w^E(.,.)$
\begin{eqnarray}
P^{[2.5]}_T \left[ P(.)  ; w^E(.,.) \right] \opsimeq_{T \to +\infty} 
C_{2.5}\left[ P(.)  ; w^E(.,.) \right] e^{- T I_{2.5}\left[ P(.)  ; w^E(.,.) \right] }
\label{level2.5infer}
\end{eqnarray}
where $C_{2.5}\left[ P(.)  ; w^E(.,.) \right]$
summarizes the constitutive constraints 
Eqs \ref{rho1ptnormaj} \ref{wdiagempi} \ref{contrainteqinfer}
\begin{eqnarray}
C_{2.5}\left[ P(.)  ; w^E(.,.) \right]
  = \delta \left( \sum_x  P(x) - 1 \right) 
   \left[ \prod_y \delta \left( \sum_x w^E(x,y)  \right) \right]
 \left[ \prod_x \delta \left( \sum_y w^E(x,y) P(y) \right) \right]
\label{constraints2.5infer}
\end{eqnarray}
while the rate function at Level 2.5 translated from Eq. \ref{rate2.5master} reads
\begin{eqnarray}
I_{2.5}\left[ P(.)  ; w^E(.,.) \right]
&&  = \sum_{y } P(y)  \sum_{x \ne y} 
\left[ w^E(x,y)   \ln \left( \frac{ w^E(x,y)   }{  w(x,y)   }  \right) 
 - w^E(x,y)  + w(x,y)   \right] 
\label{rate2.5infer}
\end{eqnarray}

So Eq. \ref{level2.5infer} describes how the empirical probability $P(.)$ and the empirical generator $w^E(.,.)$
seen on a large time-window can fluctuate around the steady state $P_*(.)$ and the true generator $w(.,.)$.


\subsection{ Generating function of the empirical probability $P(.)$ and the empirical flows $Q(.,.)$  }  

Instead of characterizing the joint statistics 
of the empirical probability $P(.)$ and of the empirical flows $Q(.,.)$
via their joint distribution $P^{[2.5]}_T \left[ P(.)  ; Q(.,.) \right] $ of Eq. \ref{level2.5master},
one can consider
their joint generating function $Z_T^{[\omega(.); \lambda(.,.)]}(x \vert x_0) $ 
over the Markov trajectories $x(t_0 \leq s \leq t) $ starting at $x(t_0)=x_0$ and ending at $x(t)=x$,
where the function $\omega(.) $ is conjugated to the empirical probability $P(.)$,
while the function $\lambda(.,.) $ is conjugated to the empirical flows $Q(.,.)$
\begin{eqnarray}
Z_T^{[\omega(.); \lambda(.,.)]}(x \vert x_0) 
&& \equiv \langle \delta_{x(T),x} \ e^{ \displaystyle T \left[ 
 \sum_{y  } P(y)  \omega( y)  +       \sum_{y  }\sum_{ x \ne y  }Q(x,y)   \lambda(x , y)  
\right] }  \ \delta_{x(0),x_0} \rangle
\nonumber \\
&& =  
 \langle \delta_{x(T),x} \ e^{ \displaystyle  \int_0^T dt  \omega( x(t)) +  
   \sum_{t \in [0,T] : x(t^+) \ne x(t^-) } \lambda(x(t^+) , x(t^-)  )  }  \ \delta_{x(0),x_0} \rangle
\label{geneddef}
\end{eqnarray}


\subsubsection{ Consequences of the constraints on the empirical observables $[P(.);Q(.,.)]$ for their conjugated variables $ [\omega(.); \lambda(.,.)]$  } 

\label{subsec_conjugc2.5}

For later purposes, it is important to stress here the consequences 
of the constitutive constraints on the empirical observables $[P(.);Q(.,.)]$ 
for their conjugated variables $ [\omega(.); \lambda(.,.)]$ :

(a) the normalization of Eq. \ref{rho1ptnormaj} for the empirical probability $P(.)$
means that only the inhomogeneities of the function $\omega(.)$ are relevant.
Indeed, if one adds an arbitrary constant $c$ to the function $\omega(x)$ for all configurations $x$
\begin{eqnarray}
 \omega_c(x) \equiv \omega(x) +c
\label{omegaavecconstant}
\end{eqnarray}
then the corresponding contribution in the exponential of the generating function of Eq. \ref{geneddef} will be 
 only shifted by a constant term
\begin{eqnarray}
 \sum_{x  } P(x) \omega_c(x)=   \sum_{x  } P(x)  \omega( x) +  c
\label{omegaavecconstantexp}
\end{eqnarray}

(b) the constitutive stationarity constraints of Eq. \ref{contrainteq} for the empirical flows $Q(.,.)$
yields that if one adds the difference $[\phi(x)-\phi(y)]$ of an arbitrary function $\phi(.)$ to the function $\lambda(x , y)$
\begin{eqnarray}
 \lambda_{[\phi(.)]}(x,y) \equiv \lambda(x , y)   + \phi(x)-\phi(y)
\label{lambdabydifference}
\end{eqnarray}
then the corresponding contribution in the exponential of the generating function of Eq. \ref{geneddef}
is unchanged using changes of label to apply Eq. \ref{contrainteq}
\begin{eqnarray}
   \sum_{y  }\sum_{ x \ne y  }Q(x,y)   \lambda_{[\phi(.)]}(x,y)
   && = \sum_{y  }\sum_{ x \ne y  }Q(x,y)   \lambda(x , y) 
   + \sum_{y  }\sum_{ x \ne y  }Q(x,y)  \phi(x)
   - \sum_{y  }\sum_{ x \ne y  }Q(x,y) \phi(y) 
   \nonumber \\
   && = \sum_{y  }\sum_{ x \ne y  }Q(x,y)   \lambda(x , y) 
   \label{lambdabydifferenceexp}
\end{eqnarray}


\subsubsection{ Eigenvalue problem governing the generating function $Z_T^{[\omega(.); \lambda(.,.)]}(x \vert x_0)  $ for large $T$  } 

\label{subsec_eigenpbZ}

For large $T$, the leading behavior of the generating function of Eq. \ref{geneddef}
\begin{eqnarray}
Z_T^{[\omega(.); \lambda(.,.)]}(x \vert x_0) 
  \opsimeq_{T \to + \infty} 
 e^{T G[\omega(.); \lambda(.,.)] } \ \ r^{[\omega(.); \lambda(.,.)]}(x) \ \  l^{[\omega(.); \lambda(.,.)]}(x_0)
\label{geneddomin0}
\end{eqnarray}
involves the highest eigenvalue $G[\omega(.); \lambda(.,.)]$ of the following deformed-matrix
$w^{[\omega(.); \lambda(.,.)]} $ with respect to the initial Markov matrix $w(.,.)$ of Eq. \ref{mastereq} 
\begin{eqnarray}
w^{[\omega(.); \lambda(.,.)]}(x, y) && \equiv w(x , y) e^{ \lambda(x,y) } \ \ \ \ \ \ \ \ \ {\rm for } \ \ x \ne y
\nonumber \\
w^{[\omega(.); \lambda(.,.)]}(x, x) && \equiv \omega(x)+w(x,x)
\label{markovmatrixdeformed}
\end{eqnarray}
The positive left eigenvector $l^{[\omega(.); \lambda(.,.)]}(x)  \geq 0$ 
\begin{eqnarray}
G[\omega(.); \lambda(.,.)]   l^{[\omega(.); \lambda(.,.)]}(  y)   
&& =\sum_x  l^{[\omega(.); \lambda(.,.)]}(  x) w^{[\omega(.); \lambda(.,.)]}(x, y)
\nonumber \\
&& = l^{[\omega(.); \lambda(.,.)]}(  y) \left[ \omega(y)+w(y,y)\right]
+ \sum_{x \ne y}  l^{[\omega(.); \lambda(.,.)]}(  x) 
w(x , y) e^{ \lambda(x,y) }
 \label{eigenleftwdef}
\end{eqnarray}
corresponds to the deformation of the trivial eigenvector $l(x)=1$ of Eq. \ref{markovleft},
while the corresponding positive right 
eigenvector $ r^{[\omega(.); \lambda(.,.)]}(x)   \geq 0 $ 
corresponds to the deformation of the right eigenvector $r(x)=P_*$ of Eq. \ref{markovright}
\begin{eqnarray}
G[\omega(.); \lambda(.,.)]    r^{[\omega(.); \lambda(.,.)]}(  x) 
&&  = \sum_y w^{[\omega(.); \lambda(.,.)]}(x, y)r^{[\omega(.); \lambda(.,.)]}(  y) 
\nonumber \\
&& =   \left[  \omega(x)+w(x,x)\right] r^{[\omega(.); \lambda(.,.)]}(  x) 
+  \sum_{y \ne x} w(x , y) e^{ \lambda(x,y) } r^{[\omega(.); \lambda(.,.)]}(  y) 
\label{eigenrightwdef}
\end{eqnarray}
with the normalization
\begin{eqnarray}
1 = \langle l^{[\omega(.); \lambda(.,.)]}  \vert r^{[\omega(.); \lambda(.,.)]} \rangle 
= \sum_x l^{[\omega(.); \lambda(.,.)]}(x)  r^{[\omega(.); \lambda(.,.)]}(x) 
 \label{Wknorma}
\end{eqnarray}



\subsubsection{ Functional Legendre transform between the explicit rate function $ I_{2.5}  [ P(.),  Q(.,.)]  $ 
and the eigenvalues $G[\omega(.);\lambda(.,.)]  $  } 

The explicit rate function $ I_{2.5}  [ P(.),  Q(.,.)]  $ of Eq. \ref{rate2.5master} governing the joint probability distribution
$P^{[2.5]}_T \left[ P(.)  ; Q(.,.) \right] $ of Eq. \ref{level2.5master}
and the (usually non-explicit) eigenvalues $G[\omega(.);\lambda(.,.)]  $ governing the generating function $Z_T^{[\omega(.); \lambda(.,.)]}(x \vert x_0)  $ via Eq. \ref{geneddomin0} are related via the following functional Legendre transformations \cite{chetrite_formal}:

(i) The generating function $Z_T^{[\omega(.); \lambda(.,.)]}(x \vert x_0) $ of Eq. \ref{geneddef}
can be computed from the joint probability distribution
$P^{[2.5]}_T \left[ P(.)  ; Q(.,.) \right] $ of Eq. \ref{level2.5master}
\begin{eqnarray}
&& Z_T^{[\omega(.); \lambda(.,.)]}  = \int dP(.)  \int d Q(.,.)P^{[2.5]}_T \left[ P(.)  ; Q(.,.) \right]
e^{ \displaystyle T \left[ 
 \sum_{y  } P(y)  \omega( y)  +       \sum_{y  }\sum_{ x \ne y  }Q(x,y)   \lambda(x , y)  
\right] }
\nonumber \\
&&  \opsimeq_{T \to + \infty} 
\int dP(.)  \int d Q(.,.) C_{2.5} \left[ P(.)  ; Q(.,.) \right]
e^{ \displaystyle T \left[ 
 \sum_{y  } P(y)  \omega( y)  +       \sum_{y  }\sum_{ x \ne y  }Q(x,y)   \lambda(x , y)  - I_{2.5}\left[ P(.)  ; Q(.,.) \right] 
\right] }
 \nonumber \\ &&
 \opsimeq_{T \to + \infty}  
 e^{ \displaystyle T G[\omega(.); \lambda(.,.)] }
\label{geneddomin0saddle}
\end{eqnarray}
via the saddle-point method for large $T$ : so $G[\omega(.); \lambda(.,.)] $
corresponds to the optimal value of the function in the exponential over the empirical probability $P(.)$ and the empirical flows $Q(.,.)$ satisfying the constitutive constraints $ C_{2.5} \left[ P(.)  ; Q(.,.) \right]$ of Eq. \ref{constraints2.5master}
\begin{eqnarray}
G[\omega(.); \lambda(.,.)]
= \max_{\substack{P(.)  \text{ and }  Q(.,.)\\ \text{satisfying } C_{2.5} \left[ P(.)  ; Q(.,.) \right]}}
 \left[ 
 \sum_{y  } P(y)  \omega( y)  +       \sum_{y  }\sum_{ x \ne y  }Q(x,y)   \lambda(x , y)  - I_{2.5}\left[ P(.)  ; Q(.,.) \right] 
\right] 
\label{legendrereci}
\end{eqnarray}
This optimization problem using the explicit expression of $ I_{2.5}\left[ P(.)  ; Q(.,.) \right]$ of Eq. \ref{rate2.5master}
and Lagrange multipliers to impose the constitutive constraints $C_{2.5} \left[ P(.)  ; Q(.,.) \right]$ of Eq. \ref{constraints2.5master}
leads to the eigenvalue problem for $G[\omega(.); \lambda(.,.)] $ described in the previous subsection \ref{subsec_eigenpbZ}.
Note that this eigenvalue problem is usually not exactly solvable,
so that eigenvalue $G[\omega(.); \lambda(.,.)] $ is usually not known explicitly 
in terms of the two arbitrary functions $[\omega(.); \lambda(.,.)]  $.

(ii) The reciprocal Legendre transformation of Eq. \ref{legendre} 
\begin{eqnarray}
I_{2.5}\left[ P(.)  ; Q(.,.) \right]
= \max_{\omega(.)  \text{ and }  \lambda(.,.)}
 \left[ 
 \sum_{y  } P(y)  \omega( y)  +       \sum_{y  }\sum_{ x \ne y  }Q(x,y)   \lambda(x , y)  -  G[\omega(.); \lambda(.,.)]
\right] 
\label{legendre}
\end{eqnarray}
leads to the explicit expression of $ I_{2.5}\left[ P(.)  ; Q(.,.) \right]$ of Eq. \ref{rate2.5master}
using only that $G[\omega(.); \lambda(.,.)] $ is the eigenvalue for the eigenvalue problem for $G[\omega(.); \lambda(.,.)] $ described in the previous subsection \ref{subsec_eigenpbZ} (since  
the eigenvalues $G[\omega(.); \lambda(.,.)] $ are not known explicitly in terms of $[\omega(.); \lambda(.,.)]  $ in most applications).



\subsection{ Canonical conditionings with respect to the empirical probability $P(.)$ and to the empirical flows $Q(.,.)$  }

\subsubsection{ Conditioned Markov generator $w^{Cond[\omega(.); \lambda(.,.)]}(.,.) $ associated to the deformed-generator $w^{[\omega(.); \lambda(.,.)]}(.,.) $} 

\label{subsec_directcond}

Since the generating function $Z_T^{[\omega(.); \lambda(.,.)]}(x \vert x_0)$ of Eq. \ref{geneddomin0}
will grow exponentially in time if the eigenvalue is positive $G[\omega(.); \lambda(.,.)] >0$
or decay exponentially in time if the eigenvalue is negative $G[\omega(.); \lambda(.,.)] <0$,
it is useful to construct the conditioned propagator (see the two very detailed papers \cite{chetrite_conditioned,chetrite_optimal} and references therein)
\begin{eqnarray}
P_T^{Cond[\omega(.); \lambda(.,.)]}(x \vert x_0) && \equiv e^{- T G[\omega(.); \lambda(.,.)] } \frac{ l^{[\omega(.); \lambda(.,.)]}(x)}{ l^{[\omega(.); \lambda(.,.)]}(x_0)} Z_T^{[\omega(.); \lambda(.,.)]}(x \vert x_0) 
\nonumber \\
&&  \opsimeq_{T \to + \infty} 
 l^{[\omega(.); \lambda(.,.)]}(x) \ \  r^{[\omega(.); \lambda(.,.)]}(x) \equiv
P^{Cond[\omega(.); \lambda(.,.)]}_* (x)
\label{conditionedpropagator}
\end{eqnarray}
that converges, independently of the initial condition $x_0$,
 towards the conditioned steady state $P^{Cond[\omega(.); \lambda(.,.)]}_* (x) $
given by the product of the left eigenvector $l^{[\omega(.); \lambda(.,.)]}(x) $ of Eqs  \ref{eigenleftwdef}
and the right eigenvector $r^{[\omega(.); \lambda(.,.)]}(x) $ of Eq \ref{eigenrightwdef}.

The corresponding conditioned Markov generator $w^{Cond[\omega(.); \lambda(.,.)]}(.,.) $ 
can be constructed as follows.
The off-diagonal element $w^{Cond[\omega(.); \lambda(.,.)]}(x,y)$ for $x \ne y$
\begin{eqnarray}
{\rm for } \ \ x \ne y : \ \ w^{Cond[\omega(.); \lambda(.,.)]}(x,y) 
&& =   l^{[\omega(.); \lambda(.,.)]}(x)  w^{[\omega(.); \lambda(.,.)]}(x , y)  \frac{1}{l^{[\omega(.); \lambda(.,.)]}(y)}  
\nonumber \\
&& =   l^{[\omega(.); \lambda(.,.)]}(x)  w(x , y) e^{ \lambda(x,y) }  \frac{1}{l^{[\omega(.); \lambda(.,.)]}(y)} 
\label{wjumpforwardklargedev}
\end{eqnarray}
involves the initial off-diagonal element $w(x,y) $, the function $\lambda(x,y) $ and 
the left eigenvector $l^{[\omega(.); \lambda(.,.)]}(.) $.
The diagonal element $w^{Cond[\omega(.); \lambda(.,.)]}(y,y) $ is determined in terms of the off-diagonal elements 
$w^{Cond[\omega(.); \lambda(.,.)]}(x,y)  $ with $x \ne y$ by the conservation of probability analog to Eq. \ref{wdiag}
and can be rewritten using the left eigenvalue Eq. \ref{eigenleftwdef} 
\begin{eqnarray}
w^{Cond[\omega(.); \lambda(.,.)]}(y,y) && = - \sum_{x \ne y} w^{Cond[\omega(.); \lambda(.,.)]}(x,y) 
 = -  \left[ \sum_{x \ne y}  l^{[\omega(.); \lambda(.,.)]}(x)  w^{[\omega(.); \lambda(.,.)]}(x , y)  \right]
 \frac{1}{l^{[\omega(.); \lambda(.,.)]}(y)}
 \nonumber \\ &&
 = -  \left[ G[\omega(.); \lambda(.,.)]   l^{[\omega(.); \lambda(.,.)]}(  y) -  l^{[\omega(.); \lambda(.,.)]}(y)  w^{[\omega(.); \lambda(.,.)]}(y , y)  \right]   \frac{1}{l^{[\omega(.); \lambda(.,.)]}(y)}
\nonumber \\
&& = w^{[\omega(.); \lambda(.,.)]}(y , y) - G[\omega(.); \lambda(.,.)]
\nonumber \\
&& = w(y,y) + \omega(y)- G[\omega(.); \lambda(.,.)]
\label{wjumpforwardklargedevdiag}
\end{eqnarray}
in order to obtain the final expression that involves
the initial diagonal element $w(y,y) $, the function $\omega(y) $ and the eigenvalue $G[\omega(.); \lambda(.,.)] $.


\subsubsection{ Special cases $[\omega(.); \lambda(.,.)] $ with vanishing eigenvalue $G[\omega(.); \lambda(.,.)] =0$ and trivial left eigenvector $l^{[\omega(.); \lambda(.,.)]}(x)=1 $}

For any given function $\lambda(.,.) $, one can choose the appropriate function $\omega(x)=\omega^{[\lambda(.,.)]}( x)$
corresponding to the vanishing eigenvalue $G[\omega^{[\lambda(.,.)]}(.); \lambda(.,.)]  =0$ and 
to the trivial left eigenvector $l^{[\omega^{[\lambda(.,.)]}(.); \lambda(.,.)]}(x)=1 $
 in Eq. \ref{eigenleftwdef}
\begin{eqnarray}
  \omega^{[\lambda(.,.)]}(y) && = - w(y,y)- \sum_{x \ne y}   w(x , y) e^{ \lambda(x,y) }
 \nonumber \\
 && = \sum_{x \ne y}   w(x , y) \left( 1- e^{ \lambda(x,y) } \right)
 \label{eigenleftwdefconserv}
\end{eqnarray}
where the last expression has been obtained using Eq. \ref{wdiag}.

Then the conditioned propagator $P_T^{Cond[\omega^{[\lambda(.,.)]}(.); \lambda(.,.)]}(x \vert x_0) $ of Eq. \ref{conditionedpropagator}
coincides with the generating function $Z_T^{[\omega^{[\lambda(.,.)]}(.); \lambda(.,.)]}(x \vert x_0) $
\begin{eqnarray}
P_T^{Cond[\omega^{[\lambda(.,.)]}(.); \lambda(.,.)]}(x \vert x_0) = Z_T^{[\omega^{[\lambda(.,.)]}(.); \lambda(.,.)]}(x \vert x_0) 
\label{conditionedpropagatorconserved}
\end{eqnarray}
while
the conditioned Markov generator $w^{Cond[\omega^{[\lambda(.,.)]}(.); \lambda(.,.)]}(x,y)  $ of Eqs \ref{wjumpforwardklargedev}
and \ref{wjumpforwardklargedevdiag}
coincides with the deformed kernel $w^{[\omega^{[\lambda(.,.)]}(.); \lambda(.,.)]}(.,.) $ of Eq. \ref{markovmatrixdeformed}
\begin{eqnarray}
w^{Cond[\omega^{[\lambda(.,.)]}(.); \lambda(.,.)]}(x,y) 
&& =    w^{[\omega^{[\lambda(.,.)]}(.); \lambda(.,.)]}(x , y) 
=  w(x , y) e^{ \lambda(x,y) } 
 \ \ {\rm for } \ \ x \ne y \ \ \ 
\nonumber \\
w^{Cond[\omega^{[\lambda(.,.)]}(.); \lambda(.,.)]}(y,y) &&  = w^{[\omega^{[\lambda(.,.)]}(.); \lambda(.,.)]}(y , y) 
= - \sum_{x \ne y}   w(x , y) e^{ \lambda(x,y) }
\label{wCondconserved}
\end{eqnarray}

These special cases $[\omega^{[\lambda(.,.)]}(.); \lambda(.,.)]$ with
vanishing eigenvalues $G[\omega^{[\lambda(.,.)]}(.); \lambda(.,.)]  =0$ and 
and trivial left eigenvectors $l^{[\omega^{[\lambda(.,.)]}(.); \lambda(.,.)]}(x)=1 $
will be very useful to analyze the inverse problem discussed in the next section.


\section{ Inverse problem for the conditioning of Markov jump processes  }

\label{sec_Inverse}

In the subsection  \ref{subsec_directcond} of the previous section,
we have recalled the direct conditioning problem,
 where a given Markov kernel $w(.,.)$ 
is conditioned with respect to some functions $ [\omega(.);\lambda(.,.)]$
in order to construct the canonical conditioned 
Markov kernel $w^{Cond[\omega(.);\lambda(.,.)]}(.,.) $ given by 
Eqs \ref{wjumpforwardklargedev} and \ref{wjumpforwardklargedevdiag}.

The goal of the present section is to analyze the following inverse problem:
 another Markov kernel ${\mathring w}(.,.)$ is given in the same configuration space as the kernel $w(.,.)$,
and one  wishes to determine whether it is possible to interpret ${\mathring w}(.,.)$
as the canonical conditioned Markov kernel $w^{Cond[\omega(.);\lambda(.,.)]}(.,.) $
 given by Eqs \ref{wjumpforwardklargedev} and \ref{wjumpforwardklargedevdiag}
\begin{eqnarray}
{\mathring w}(x,y) && = w^{Cond[\omega(.); \lambda(.,.)]}(x,y) 
 =   l^{[\omega(.); \lambda(.,.)]}(x)  w(x , y) e^{ \lambda(x,y) }  \frac{1}{l^{[\omega(.); \lambda(.,.)]}(y)} 
 \ \ {\rm for } \ \ x \ne y \ \ \ 
\nonumber \\
{\mathring w}(y,y) && = w^{Cond[\omega(.); \lambda(.,.)]}(y,y) = w(y,y) + \omega(y)- G[\omega(.); \lambda(.,.)]
\label{winverse}
\end{eqnarray}
with some appropriate functions $ [\omega(.);\lambda(.,.)]$ that have to be determined.

Since the conditioned Markov kernel $w^{Cond[\omega(.);\lambda(.,.)]}(.,.) $
only changes the values of the non-vanishing matrix elements of $w(.,.)$,
but cannot create new transitions that do not exist in $w(.,.)$,
an obvious condition for the inverse problem is that the two given Markov kernels ${\mathring w}(.,.)  $ and $w(.,.)  $
 should involve the same possible transitions in configuration space: for any pair of configurations $x \ne y$, 
the transition from configuration $y$ towards  $x$ is possible 
for the new dynamics ${\mathring w}(x,y) > 0  $  
only if this transition from $y$ to $x$ is already possible in the initial dynamics $w(x,y) > 0 $.
Once this condition is fulfilled, the goal is to determine 
all the possible functions $ [\omega(.);\lambda(.,.)]$
satisfying Eq. \ref{winverse} in terms of the two given Markov kernels $ {\mathring w}(.,.) $ and $w(.,.)$.


\subsection{ Simple solution $[{\mathring \omega}(.); {\mathring \lambda}(.,.)]$ 
with vanishing eigenvalue $G[{\mathring \omega}(.); {\mathring \lambda}(.,.)] =0$
and left eigenvector $l^{[{\mathring \omega}(.); {\mathring \lambda}(.,.)]}(x)=1 $ } 

If one chooses ${\mathring \omega} (y)=\omega^{[\mathring \lambda(.,.)]}(y)$ of Eq. \ref{eigenleftwdefconserv}
corresponding to the vanishing eigenvalue $G[{\mathring \omega}(.); {\mathring \lambda}(.,.)] =0$ and 
to the trivial left eigenvector $l^{[{\mathring \omega}(.); {\mathring \lambda}(.,.)]}(x)=1 $,
then the inverse problem of Eq. \ref{winverse}
reduces to
\begin{eqnarray}
{\mathring w}(x,y) &&  =   w(x , y) e^{ {\mathring \lambda}(x,y) } 
 \ \ {\rm for } \ \ x \ne y \ \ \ 
\nonumber \\
{\mathring w}(y,y) && =  w(y,y) + {\mathring \omega}(y)
\label{winverseconserved}
\end{eqnarray}
and leads to the simple solution
\begin{eqnarray}
{\mathring \lambda}(x,y) && = \ln \left( \frac{ {\mathring w}(x,y) }{  w(x , y) } \right)
 \ \ {\rm for } \ \ x \ne y \ \ \ 
 \nonumber \\
 {\mathring \omega}(y) && = {\mathring w}(y,y) - w(y , y)
\label{wCondconservedsolinverse}
\end{eqnarray}

This simple solution $[{\mathring \omega}(.); {\mathring \lambda}(.,.)]$ can be also directly understood 
via the rewriting of the trajectory probability ${\cal P }_{[{\mathring w}(.,.)]}\left[ x(0 \leq t \leq T) \right]  $ 
associated to the Markov generator ${\mathring w}(.,.) $
in terms of the trajectory probability 
${\cal P }_{[w(.,.)]}\left[ x(0 \leq t \leq T) \right] $ 
associated to the initial Markov generator $w(.,.) $
as given by Eq. \ref{pwtrajjump} in terms of the empirical observables $P(.);Q(.,.)] $ associated to the trajectory $x(0 \leq t \leq T) $
\begin{eqnarray}
&&{\cal P }_{[{\mathring w}(.,.)]}\left[ x(0 \leq t \leq T) \right]  
 \equiv e^{ \displaystyle T \left[ \sum_x P(x){\mathring w}(x,x)  + \sum_y \sum_{x \ne y}  
  Q(x,y) \ln ( {\mathring w}(x,y ) ) \right] }
 \nonumber \\
 && = e^{ \displaystyle T \left[ \sum_x P(x)w(x,x)  + \sum_y \sum_{x \ne y}  
  Q(x,y) \ln ( w(x,y ) ) \right] }
 \times e^{ \displaystyle T \left[ \sum_x P(x) \left[ {\mathring w}(x,x) - w(x,x)\right] + \sum_y \sum_{x \ne y}  
  Q(x,y)  \ln \left( \frac{ {\mathring w}(x,y) }{  w(x , y) } \right) \right] }
  \nonumber \\
 && = {\cal P }_{[w(.,.)]}\left[ x(0 \leq t \leq T) \right] 
 \times e^{ \displaystyle T \left[ \sum_x P(x) {\mathring \omega}(x) + \sum_y \sum_{x \ne y}  
  Q(x,y) {\mathring \lambda}(x,y)  \right] }
\label{pwtrajjumpring}
\end{eqnarray}


\subsection{ Other solutions $ [\omega(.);\lambda(.,.)]$  corresponding to other eigenvalues $G$ and other eigenvectors $l(.)$ } 

The other solutions $ [\omega(.);\lambda(.,.)]$  for the inverse problem of Eq. \ref{winverse}
can be found using the simple solution $[{\mathring \omega}(.); {\mathring \lambda}(.,.)] $ of Eq. \ref{winverseconserved} 
that can be plugged into Eq. \ref{winverse} to obtain the equations
\begin{eqnarray}
w(x , y) e^{ {\mathring \lambda}(x,y) } && =    l^{[\omega(.); \lambda(.,.)]}(x)  w(x , y) e^{ \lambda(x,y) }  \frac{1}{l^{[\omega(.); \lambda(.,.)]}(y)} 
 \ \ {\rm for } \ \ x \ne y \ \ \ 
\nonumber \\
w(y,y) + {\mathring \omega}(y) && = w(y,y) + \omega(y)- G[\omega(.); \lambda(.,.)]
\label{winversering}
\end{eqnarray}
So the function $\omega(y) $ can differ from the simple solution ${\mathring \omega}(y) $
only by the constant corresponding to the eigenvalue $G[\omega(.); \lambda(.,.)] $
\begin{eqnarray}
\omega(y) = {\mathring \omega}(y) + G[\omega(.); \lambda(.,.)]
\label{omegaomegaring}
\end{eqnarray}
while the function $\lambda(x,y) $ can differ from the simple solution ${\mathring \lambda}(x,y) $ only 
via some difference of the logarithm of the left eigenvector $ l^{[\omega(.); \lambda(.,.)]}(.) $
\begin{eqnarray}
\lambda(x,y) =  {\mathring \lambda}(x,y)  - \ln \left( l^{[\omega(.); \lambda(.,.)]}(x)  \right) + \ln \left( l^{[\omega(.); \lambda(.,.)]}(y)  \right)
 \ \ {\rm for } \ \ x \ne y \ \ \ 
\label{lambdalambdaring}
\end{eqnarray}
Note that the relations of Eqs \ref{omegaomegaring} and \ref{lambdalambdaring} between the different solutions
can be understood from the properties of $ [\omega(.); \lambda(.,.)]$ discussed in subsection \ref{subsec_conjugc2.5}.

In summary, the choice of the eigenvalue $G$ determines the function $\omega(y) $ via Eq. \ref{omegaomegaring}
\begin{eqnarray}
\omega^{[G]}(y) = {\mathring \omega}(y) + G= {\mathring w}(y,y) - w(y , y)+G
\label{omegaomegaringsolugene}
\end{eqnarray}
while the choice of the positive left eigenvector $l(.)>0$ determines the function 
$\lambda(x,y) $ via Eq. \ref{lambdalambdaring}
\begin{eqnarray}
\lambda^{[l(.)]}(x,y) =  {\mathring \lambda}(x,y)  - \ln \left( l(x)  \right) + \ln \left( l(y)  \right)
= \ln \left( \frac{ {\mathring w}(x,y) }{  w(x , y) } \right)- \ln \left( l(x)  \right) + \ln \left( l(y)  \right)
 \label{lambdalambdaringsolugene}
\end{eqnarray}
In concrete applications, one can take advantage of this freedom to choose the eigenvalue $G$ 
and the positive left eigenvector $l(.)>0$ in order to simplify the two functions $ [\omega(.); \lambda(.,.)]$
and to have a simple physical meaning for the corresponding conditioning observable.

In the remaining sections of the main text, this general solution for the inverse problem
is illustrated with the simple examples of
a single particle on a one-dimensional periodic ring with directed dynamics (Section \ref{sec_Directed}) or undirected dynamics (Section \ref{sec_UnDirected}), before we turn to
many-body spin models with single-spin dynamics (Section \ref{sec_SingleSpinFlip}) or two-spin-flip dynamics (Section \ref{sec_TwoSpinFlip}).


\section{ Application to Directed Markov jump processes on a periodic ring } 

\label{sec_Directed}

The Directed Markov jump processes on a one-dimensional periodic ring
are the simplest non-equilibrium Markov jump processes 
(the directed character of the dynamics prevents any possibility of detailed-balance),
whose large deviations properties have been analyzed recently 
from various perspectives \cite{c_ring,c_ruelle,c_inference}.
The most general Directed Trap model 
on the ring of $L$ sites with periodic boundary conditions $x+L \equiv x$
involves arbitrary trapping times $\tau_x$ on the $L$ sites $x=1,2,..,L$ 
\begin{eqnarray}
{\mathring w}(x+1,x) &&  =  \frac{1}{\tau_x}
\nonumber \\
{\mathring w}(x,x) && =  - \frac{1}{\tau_x}
\label{DirectedTrap}
\end{eqnarray}
The simplest dynamics with the same possible transitions is the
uniform Directed Trap model where all the trapping times are instead unity 
\begin{eqnarray}
w(x+1,x) &&  =  1
\nonumber \\
w(x,x) && =  - 1
\label{DirectedTrapuniform}
\end{eqnarray}
so that the corresponding steady state is uniform $P_*(x)=\frac{1}{L}$ 
and with the uniform steady flow $Q_*(x+1,x)= \frac{1}{L}$ along the ring.

Let us first discuss the direct problem of the canonical conditioning 
of the pure directed trap model of Markov generator $w$ 
of Eq. \ref{DirectedTrapuniform}
before considering the inverse problem where one wishes to interpret ${\mathring w} $ of Eq. \ref{DirectedTrap}
as some canonical conditioning of $w(.,.)$.


\subsection{ Direct problem : canonical conditioning of 
the generator $w$ with arbitrary functions $[\omega(.); \lambda(.,.) ]$} 

In the canonical conditioning of the generator $w(.,.)$ of Eq. \ref{DirectedTrapuniform},
the function $\omega(x)$ is conjugated to the normalized empirical density $P(x)$,
while the function $\lambda(x) \equiv \lambda(x+1,x)$ is conjugated to the empirical flow $Q(x+1,x)$
that should be uniform along the ring in order to satisfy the constitutive constraint of Eq. \ref{contrainteq}
\begin{eqnarray}
Q(y+1,y)=Q
 \label{Qconserved}
\end{eqnarray}
As a consequence, the conditioning observable
\begin{eqnarray}
\sum_{x=0}^{L-1} \left[ \omega(x) P(x) + \lambda(x) Q(x+1,x) \right]=
\sum_{x=0}^{L-1}  \omega(x) P(x) + Q  \lambda^{tot}
\ \ \ \text{ with } \ \ \lambda^{tot} \equiv  \sum_{x=0}^{L-1} \lambda(x)   
 \label{conserflowalongthering}
\end{eqnarray}
involves the function $\omega(x)$ and the global parameter $\lambda^{tot}$
conjugated to the conserved flow $Q$.

The eigenvalue Eq. \ref{eigenleftwdef} 
for the positive left eigenvector $l^{[\omega(.); \lambda(.)]}(  .) >0$
when the generator $w(.,.)$ is given by Eq. \ref{DirectedTrapuniform} 
reduces to the simple recursion 
\begin{eqnarray}
 l^{[\omega(.); \lambda(.)]}(  y+1)  
 =e^{ - \lambda(y) }\left( G[\omega(.); \lambda(.)]  +1-\omega(y) \right)  l^{[\omega(.); \lambda(.)]}(  y)     
 \label{eigenleftwdeftrap}
\end{eqnarray}
 so all the coefficients should be positive $ \left(G[\omega(.); \lambda(.)]  +1-\omega(y) \right)>0$.
In addition, the left eigenvector should satisfy the periodic boundary conditions $ l^{[\omega(.); \lambda(.)]}(  y=L)=l^{[\omega(.); \lambda(.)]}(  y=0)$, so that
the eigenvalue $G[\omega(.); \lambda(.)] $ satisfies the closed equation
\begin{eqnarray}
1 && =  \prod_{y=0}^{L-1} \left[ e^{ - \lambda(y) } \left( G[\omega(.); \lambda(.)]  +1-\omega(y) \right) \right]
= e^{ - \lambda^{tot}} \prod_{y=0}^{L-1} \left[   G[\omega(.); \lambda(.)]  +1-\omega(y)  \right]
 \label{eigenleftwdeftrapsolG}
\end{eqnarray}
that depends only on the $\lambda(x)$ via the global parameter $\lambda^{tot}$ of Eq. \ref{conserflowalongthering}.

The off-diagonal elements of the conditioned generator of Eq. \ref{wjumpforwardklargedev} read using Eq. \ref{eigenleftwdeftrap}
\begin{eqnarray}
w^{Cond[\omega(.); \lambda(.,.)]}(y+1,y) 
 =      e^{ \lambda(y) }  \frac{l^{[\omega(.); \lambda(.)]}(y+1)}{l^{[\omega(.); \lambda(.)]}(y)} 
 = G[\omega(.); \lambda(.)]  +1-\omega(y)
\label{wjumpforwardklargedevtrap}
\end{eqnarray}
while the diagonal elements of Eq. \ref{wjumpforwardklargedevdiag} 
\begin{eqnarray}
w^{Cond[\omega(.); \lambda(.)}(y,y)  =  \omega(y)-1- G[\omega(.); \lambda(.)]
\label{wjumpforwardklargedevdiagtrap}
\end{eqnarray}
are consistent with the conservation of probability $w^{Cond[\omega(.); \lambda(.)}(y,y)
= -  w^{Cond[\omega(.); \lambda(.,.)]}(y+1,y) $ as it should.

In summary, to solve the direct problem of the canonical conditioning of 
the Markov generator $w$ of Eq. \ref{DirectedTrapuniform}
with arbitrary functions $[\omega(.); \lambda(.,.) ]$,
one needs to compute the eigenvalue $G[\omega(.); \lambda(.)] $
from the closed Eq. \ref{eigenleftwdeftrapsolG}
and to plug it into Eqs \ref{wjumpforwardklargedevtrap} and \ref{wjumpforwardklargedevdiagtrap} to obtain the conditioned generator $w^{Cond[\omega(.); \lambda(.)}(.,.) $.


\subsection{ Inverse problem : functions $[\omega(x),\lambda(x)]$ to produce 
the given generator ${\mathring w}(.,.)=w^{Cond[\omega(.); \lambda(.,.)]}(.,.)$} 

The general solution discussed in section \ref{sec_Inverse}
can be applied to the two given generators of Eqs \ref{DirectedTrap}
and \ref{DirectedTrapuniform} as follows:

(i) the simple solution of Eq. \ref{wCondconservedsolinverse} corresponding
to the vanishing eigenvalue $G[{\mathring \omega}(.); {\mathring \lambda}(.)] =0$
and 
to the trivial left eigenvector $l^{[{\mathring \omega}(.); {\mathring \lambda}(.)] }(x)=1 $
reads
\begin{eqnarray}
{\mathring \lambda}(x) && = \ln \left( \frac{ {\mathring w}(x+1,x) }{  w(x+1 , x) } \right) =   \ln \left( \frac{1}{\tau_x} \right)
 \nonumber \\
 {\mathring \omega}(x) && = {\mathring w}(x,x) - w(x , x) = 1 - \frac{1}{\tau_x}
\label{wCondconservedsolinversetrap}
\end{eqnarray}

(ii) the other solutions can be parametrized by the eigenvalue $G$ that determines 
the function $\omega(x) $ via Eq. \ref{omegaomegaringsolugene}
\begin{eqnarray}
\omega^{[G]}(x)  =  (1+G) - \frac{1}{\tau_x}
\label{omegaxtrapG}
\end{eqnarray}
and by the positive left eigenvector $l(.)>0$ that determines the function 
$\lambda(x)=\lambda(x+1,x) $ via Eq. \ref{lambdalambdaringsolugene}
\begin{eqnarray}
\lambda^{[l(.)]}(x) =   \ln \left( \frac{1}{\tau_x} \right)  - \ln \left( l(x+1)  \right) + \ln \left( l(x)  \right)
 \label{lambdalambdaringsolugenetrap}
\end{eqnarray}
Since the sum along the ring cannot be changed by the choice of the periodic
eigenvector $l(.)$
\begin{eqnarray}
\sum_{x=0}^{L-1} \lambda^{[l(.)]}(x) = \sum_{x=0}^{L-1}  \ln \left( \frac{1}{\tau_x} \right)  \equiv \lambda^{tot}
 \label{lambdalambdaringsolugenetrapsum}
\end{eqnarray}
the parameter $\lambda^{tot} $ of Eq. \ref{conserflowalongthering}
conjugated to the conserved empirical flow $Q(y+1,y)=Q$ along the ring is fixed by 
the sum of the logarithms of the trapping times $\tau_x$ along the ring.
To be more concrete, let us now describe two examples.


\subsubsection{ First example when the kernel ${\mathring w}(x+1,x)=- {\mathring w}(x,x)= \frac{1}{\tau} $ is translation-invariant along the ring   }

When all the trapping times of Eq. \ref{DirectedTrap} coincide $\tau_x=\tau$,
then the kernel ${\mathring w}(x+1,x)=- {\mathring w}(x,x) =\frac{1}{\tau} $ is translation-invariant along the ring.
One can then choose the value $G=\frac{1}{\tau}-1 $ to make Eq. \ref{omegaxtrapG} vanish
\begin{eqnarray}
\text{Choice : } \ \ \ \omega^{[G=\frac{1}{\tau}-1 ]}(x)  =  0
\label{omegaxtrapGzero}
\end{eqnarray}
Then there is no conditioning with respect to the empirical density $P(y)$ but only with respect to the 
conserved empirical flow $Q(x+1,x)=Q$ along the ring via the conjugated parameter $\lambda^{tot}$
fixed by Eq. \ref{lambdalambdaringsolugenetrapsum}, so that the conditioning observable of Eq. \ref{conserflowalongthering}
reduces to
\begin{eqnarray}
\sum_{x=0}^{L-1} \left[ \omega(x) P(x) + \lambda(x) Q(x+1,x) \right]=
 Q  \lambda^{tot} = Q L \ln \left( \frac{1}{\tau} \right)
 \label{obstranslinvariant}
\end{eqnarray}
The conclusion is thus that the translation-invariant kernel ${\mathring w}(x+1,x)=- {\mathring w}(x,x)=\frac{1}{\tau}$  can be interpreted as the conditioning of the kernel $w$ of Eq. \ref{DirectedTrapuniform} with respect to the empirical flow $Q$ along the ring with the conjugated parameter $L \ln \left( \frac{1}{\tau} \right) $.


\subsubsection{ Second example when the trapping times satisfy the global condition $\displaystyle \sum_{x=0}^{L-1}  \ln \left( \frac{1}{\tau_x} \right) =0  $  }

When the Directed trap generator ${\mathring w} $ of Eq. \ref{DirectedTrap}
is characterized by trapping times satisfying the global condition 
\begin{eqnarray}
 \sum_{x=0}^{L-1}  \ln \left( \frac{1}{\tau_x} \right)  =0
 \label{lambdalambdaringsolugenetrapsumzero}
\end{eqnarray}
then the parameter $\lambda^{tot} $ of Eq. \ref{conserflowalongthering}
conjugated to the conserved empirical flow $Q(y+1,y)=Q$ along the ring vanishes $\lambda^{tot} =0$.
 Note that one can choose the left eigenvector $l(.)$
 in order to make vanish all the parameters $\lambda^{[l(.)]}(x) $ of Eq. \ref{lambdalambdaringsolugenetrap} 
\begin{eqnarray}
\text{ Choice : } \ \ \ 0 = \lambda^{[l(.)]}(x) =   \ln \left( \frac{1}{\tau_x} \right)  - \ln \left( l(x+1)  \right) + \ln \left( l(x)  \right)
 \label{lambdalambdaringsolugenetrapchoicezero}
\end{eqnarray}
The conditioning is then only with respect to the empirical density $P(x)$ 
and one can choose $G=-1$ to simplify the function $\omega^{[G]}(x) $ of Eq. \ref{omegaxtrapG}
\begin{eqnarray}
\text{ Choice : } \omega^{[G=-1]}(x)  =   - \frac{1}{\tau_x}
\label{omegaxtrapGchoice}
\end{eqnarray}
so that the conditioning observable of Eq. \ref{conserflowalongthering}
reduces to
\begin{eqnarray}
 \sum_{x=0}^{L-1}  \omega^{[G=-1]}(x) P(x) =   -  \sum_{x=0}^{L-1} \frac{P(x)}{\tau_x}
\label{observachoice}
\end{eqnarray}
where the trapping times inverses $ \frac{1}{\tau_x}$ are conjugated to the empirical density $P(x)$.


\section{ Application to undirected Markov jump processes on a periodic ring } 

\label{sec_UnDirected}

For undirected Markov jump processes on a one-dimensional periodic ring,
 whose large deviations properties at various levels have been discussed in \cite{c_ring},
the most general dynamics on a ring of $L$ sites
 involves arbitrary rates ${\mathring w}(x \pm 1,x)={\mathring w}^{\pm}_x$ 
 towards the two nearest-neighbors $(x \pm 1)$ from each site $x=1,..,L$
\begin{eqnarray}
{\mathring w}(x\pm1,x) &&  ={\mathring w}^{\pm}_x
\nonumber \\
{\mathring w}(x,x) && =  - {\mathring w}^+_x-{\mathring w}^-_x
\label{UnDirectedTrap}
\end{eqnarray}
The simplest dynamics with the same possible transitions is the
uniform model where all jump rates towards nearest-neighbors are instead unity 
\begin{eqnarray}
w(x \pm 1,x) &&  =  1
\nonumber \\
w(x,x) && =  - 2
\label{UnDirectedTrapuniform}
\end{eqnarray}
This generator $w$ satisfies detailed-balance with respect to its uniform steady state $P_*(x)=\frac{1}{L}$.

We will first discuss the direct problem of the canonical conditioning of the equilibrium dynamics governed by $w$ of Eq. \ref{UnDirectedTrapuniform}
before considering the inverse problem where one wishes to interpret ${\mathring w} $ of Eq. \ref{UnDirectedTrap}
as some canonical conditioning of $w(.,.)$.


\subsection{ Direct problem : canonical conditioning of 
the generator $w$ with arbitrary functions $[\omega(.); \lambda(.,.) ]$ }

The constitutive constraints of Eq. \ref{contrainteq} for the empirical flows $Q(.,.)$ between nearest-neighboring sites on the ring 
yield that they can be decomposed into their symmetric and antisymmetric parts on each link $(x,x+1)$
\begin{eqnarray}
 Q(x+1,x) && = \frac{ A\left(x+\frac{1}{2}\right) + J}{2}
\nonumber \\
Q(x,x+1) && = \frac{ A\left(x+\frac{1}{2}\right) - J}{2}
 \label{QAJ}
\end{eqnarray}
where $J$ represents the uniform empirical current flowing along the ring
\begin{eqnarray}
J=Q(x+1,x)-Q(x,x+1)
 \label{JQQ}
\end{eqnarray}
while $A\left(x+\frac{1}{2}\right)$ represents the activity of the link between the two sites $(x,x+1)$
\begin{eqnarray}
A\left(x+\frac{1}{2}\right) =Q(x+1,x) +Q(x,x+1) 
 \label{AQQ}
\end{eqnarray}
As a consequence, the conditioning observable
involving the functions $\omega(x)$ and $\lambda^{\pm}(x) \equiv \lambda(x \pm 1,x)$
\begin{eqnarray}
\sum_{x=0}^{L-1} \left[ \omega(x) P(x) + \lambda^+(x) Q(x+1,x) + \lambda^-(x+1) Q(x,x+1)\right]
&& =
\sum_{x=0}^{L-1} \left[ \omega(x) P(x) 
+ A\left(x+\frac{1}{2}\right)  \frac{ \lambda^+(x) + \lambda^-(x+1)}{2} \right]
\nonumber \\
&& + J \left[\sum_{x=0}^{L-1}   \frac{ \lambda^+(x) - \lambda^-(x+1)}{2} \right]
 \label{observableundirected}
\end{eqnarray}
can be rewritten in terms of the link activities $A\left(x+\frac{1}{2}\right) $ and in term of the global current $J$.

The eigenvalue 
Eq. \ref{eigenleftwdef} for the positive periodic left eigenvector $l^{[\omega(.); \lambda^{\pm}(.)]}(  .)$
reads 
\begin{eqnarray}
0 = l^{[\omega(.); \lambda^{\pm}(.)]}(  y) \left[ \omega(y)-2 - G[\omega(.); \lambda^{\pm}(.)] \right]
+  l^{[\omega(.); \lambda^{\pm}(.)]}(  y+1)  e^{ \lambda^+(y) }
+  l^{[\omega(.); \lambda^{\pm}(.)]}(  y-1)  e^{ \lambda^-(y) }
 \label{eigenleftwdeftrapun}
\end{eqnarray}

The off-diagonal elements of the conditioned generator can be then obtained via Eq. \ref{wjumpforwardklargedev} 
\begin{eqnarray}
w^{Cond[\omega(.); \lambda^{\pm}(.)]}(y+1,y) 
&& =     e^{ \lambda^+(y) }  \frac{l^{[\omega(.); \lambda^{\pm}(.)]}(y+1) }{l^{[\omega(.); \lambda^{\pm}(.)]}(y)} 
\nonumber \\
w^{Cond[\omega(.); \lambda^{\pm}(.)]}(y-1,y) 
&& =  e^{ \lambda^-(y) }  \frac{ l^{[\omega(.); \lambda^{\pm}(.)]}(y-1)   }{l^{[\omega(.); \lambda^{\pm}(.)]}(y)} 
\label{wjumpforwardklargedevtrapun}
\end{eqnarray}
while the diagonal elements are given by
 Eq. \ref{wjumpforwardklargedevdiag} 
\begin{eqnarray}
w^{Cond[\omega(.); \lambda^{\pm}(.)}(y,y) && =  \omega(y)-2- G[\omega(.); \lambda^{\pm}(.)]
\label{wjumpforwardklargedevdiagtrapun}
\end{eqnarray}

So for arbitrary functions $[\omega(.); \lambda^{\pm}(.,.) ]$,
the conditioned generator $w^{Cond[\omega(.); \lambda^{\pm}(.)]} $
is not explicit as a consequence of the two-terms recursion of Eq. \ref{eigenleftwdeftrapun}
with arbitrary coefficients.


\subsection{ Inverse problem : functions $[\omega(x),\lambda^{\pm}(x)]$ to produce 
the given generator ${\mathring w}(.,.)=w^{Cond[\omega(.); \lambda^{\pm}(.)]}(.,.)$} 

For the inverse problem,
the general solution of section \ref{sec_Inverse}
can be applied to the two given generators of Eqs \ref{UnDirectedTrap}
and \ref{UnDirectedTrapuniform} as follows:

(i) the simple solution of Eq. \ref{wCondconservedsolinverse} corresponding
to the vanishing eigenvalue $G[{\mathring \omega}(.); {\mathring \lambda}^{\pm}(.)] =0$
and 
to the trivial left eigenvector $l^{[{\mathring \omega}(.); {\mathring \lambda}^{\pm}(.)] }(x)=1 $
reduces to
\begin{eqnarray}
{\mathring \lambda}^{\pm}(x) && = \ln \left( \frac{ {\mathring w}(x\pm1,x) }{  w(x\pm1 , x) } \right) 
=   \ln \left( {\mathring w}^{\pm}(x) \right)
 \nonumber \\
 {\mathring \omega}(x) && = {\mathring w}(x,x) - w(x , x) = 2- {\mathring w}^+_x-{\mathring w}^-_x
\label{wCondconservedsolinversetrapun}
\end{eqnarray}

(ii) the other solutions can be parametrized by the eigenvalue $G$ that determines 
the function $\omega(x) $ via Eq. \ref{omegaomegaringsolugene}
\begin{eqnarray}
\omega^{[G]}(x)  =   2+G - {\mathring w}^+_x-{\mathring w}^-_x
\label{omegaxtrapGun}
\end{eqnarray}
and by the positive periodic left eigenvector $l(.)>0$ that determines the functions 
$\lambda^{\pm}(x) = \lambda(x\pm 1,x) $ via Eq. \ref{lambdalambdaringsolugene}
\begin{eqnarray}
\lambda^{[l(.)] \pm} (x) =   \ln \left( {\mathring w}^{\pm}(x) \right)  - \ln \left( l(x \pm 1)  \right) + \ln \left( l(x)  \right)
 \label{lambdalambdaringsolugenetrapun}
\end{eqnarray}
So the two following sums along the ring cannot be changed by the choice of the periodic
eigenvector $l(.)$
\begin{eqnarray}
\sum_{x=0}^{L-1} \lambda^{[l(.)]+}(x) && = \sum_{x=0}^{L-1}  \ln \left( {\mathring w}^{+}(x) \right)  
\nonumber \\
\sum_{x=0}^{L-1} \lambda^{[l(.)]-}(x) && = \sum_{x=0}^{L-1}  \ln \left( {\mathring w}^{-}(x) \right)
 \label{lambdalambdaringsolugenetrapun2sum}
\end{eqnarray}
These two sum rules indicate via their difference and via their sum respectively
that the conditioning of Eq. \ref{observableundirected}
with respect to the global empirical current $J$,
and with respect to the total empirical activity of the ring 
\begin{eqnarray}
A^{tot}=\sum_x A\left(x+\frac{1}{2}\right) 
 \label{Atot}
\end{eqnarray}
are determined by the two products 
$\prod_{x=0}^{L-1}  {\mathring w}^{+}(x)  $ and $\prod_{x=0}^{L-1}  {\mathring w}^{-}(x)  $.
To be more concrete, let us now describe the simplest example with translation-invariance.


\subsubsection*{ Simplest example when the kernel ${\mathring w}$ is translation-invariant along the ring }

When the kernel ${\mathring w}$ of Eq. \ref{UnDirectedTrap}
is translation-invariant along the ring
\begin{eqnarray}
{\mathring w}(x\pm1,x) &&  ={\mathring w}^{\pm}
\nonumber \\
{\mathring w}(x,x) && =  - {\mathring w}^+-{\mathring w}^-
\label{UnDirectedTrapTrnalation}
\end{eqnarray}
one can choose the simple solution of Eq. \ref{wCondconservedsolinversetrapun} 
for the parameters ${\mathring \lambda}^{\pm}(x) $ that are translation-invariant
\begin{eqnarray}
{\mathring \lambda}^{\pm} && 
=   \ln \left( {\mathring w}^{\pm} \right)
\label{wCondconservedsolinversetrapunchoice}
\end{eqnarray}
and one can then choose the parameter $G={\mathring w}^++{\mathring w}^--2$ to make Eq. \ref{omegaxtrapGun}  vanish
\begin{eqnarray}
\text{ Choice : } \ \ 0=\omega^{[G={\mathring w}^++{\mathring w}^--2]}(x)  =  0
\label{omegaxtrapGunchoice}
\end{eqnarray}
Then the conditioning observable of Eq. \ref{observableundirected} does not involve the empirical density $P(.)$
but only involves the golbal empirical current $J$ and the total activity $A^{tot}$ of Eq. \ref{Atot}
 \begin{eqnarray}
&& \sum_{x=0}^{L-1} \left[ \omega(x) P(x) 
+ A\left(x+\frac{1}{2}\right)  \frac{ \lambda^+(x) + \lambda^-(x+1)}{2} \right] + J \left[\sum_{x=0}^{L-1}   \frac{ \lambda^+(x) - \lambda^-(x+1)}{2} \right]
 \nonumber \\
&& =
 A^{tot} L \frac{ {\mathring \lambda}^+ +{\mathring \lambda}^-}{2} 
 + J  L   \frac{ {\mathring \lambda}^+ - {\mathring \lambda}^-}{2} 
= A^{tot}  \frac{ \ln \left( {\mathring w}^+{\mathring w}^- \right)}{2} 
 + J  L   \frac{ \ln \left( \frac{ {\mathring w}^+}{{\mathring w}^- } \right)}{2} 
 \label{observableundirectedtranslation}
\end{eqnarray}
The conclusion is thus that the non-equilibrium translation-invariant kernel ${\mathring w} $ of Eq. \ref{UnDirectedTrapTrnalation} involving the two parameters ${\mathring w}^{\pm} $ can be interpreted as the conditioning of the equilibrium kernel $w$ of Eq. \ref{UnDirectedTrapuniform} 
with respect to the global empirical current $J$ and with respect to the total  empirical activity $A^{tot}$.


\section{ Application to single-spin-flip dynamics of classical spins models } 

\label{sec_SingleSpinFlip}

For many-body models involving $N$ classical spins $S_j=\pm$
and thus $2^N$ configurations $C=\{S_1,..S_{i-1},S_i,S_{i+1} ...,S_N\}$,
the most general single-spin-flip dynamics
involves arbitrary rates ${\mathring w} (\sigma_i^x C,C )$ 
for the flip of the spin $S_i \to -S_i$ when starting in the configuration $C=\{S_1,..S_{i-1},S_i,S_{i+1} ...,S_N\}$
towards $\sigma_i^x C =\{S_1,..S_{i-1},-S_i,S_{i+1} ...,S_N\}$
\begin{eqnarray}
{\mathring w} (\sigma_i^x C,C ) &&  =  {\mathring w}_i(C)
\nonumber \\
{\mathring w}(C,C) && =  - \sum_{i=1}^N {\mathring w}_i(C)
\label{singlespinflip}
\end{eqnarray}
The simplest dynamics with the same possible transitions
is the single-spin-flip dynamics where each spin can flip independently with rate unity
\begin{eqnarray}
w (\sigma_i^x C,C ) &&  = 1
\nonumber \\
w(C,C) && =  - N
\label{singlespinflipindep}
\end{eqnarray}
i.e. the generator $w$ written as a quantum operator using Pauli matrices $\sigma_i^x$ reduces to
\begin{eqnarray}
w   = \sum_{i=1}^N \left( \sigma_i^x -1 \right)
\label{singlespinflipindepPauli}
\end{eqnarray}
and satisfies detailed-balance with respect to its uniform steady state $P_*(C)=\frac{1}{2^N}$ over the $2^N$ configurations.

We will first discuss the direct problem of the canonical conditioning of the equilibrium dynamics governed by $w$ of Eq. \ref{singlespinflipindepPauli}
before considering the inverse problem where one wishes to interpret ${\mathring w} $ of Eq. \ref{singlespinflip}
as some canonical conditioning of $w(.,.)$.


\subsection{ Direct problem : canonical conditioning of 
the Markov generator $w$ with arbitrary functions $[\omega(.); \lambda(.,.) ]$} 

For the direct conditioning problem of the generator $w(.,.)$ of Eq. \ref{singlespinflipindep}
involving the functions $\omega(C)$ and $\lambda_i(C) \equiv \lambda(\sigma_i^x C,C)$,
one needs to solve the eigenvalue 
Eq. \ref{eigenleftwdef} for the positive left eigenvector $l^{[\omega(.); \lambda_.(.)]}(  .)$
\begin{eqnarray}
0
 = l^{[\omega(.); \lambda_.(.)]}(  C) \left[ \omega(C)-N - G[\omega(.); \lambda_.(.)]\right]
+ \sum_{i=1}^N  l^{[\omega(.); \lambda_.(.)]}( \sigma_i^x C ) 
 e^{ \lambda_i(C) }
 \label{eigenleftsinglespinflip}
\end{eqnarray}

The off-diagonal elements of the conditioned generator can be then obtained via Eq. \ref{wjumpforwardklargedev} 
\begin{eqnarray}
w^{Cond[\omega(.); \lambda_.(.)]}(\sigma_i^x C,C) 
&& =    e^{ \lambda_i(C) }  \frac{ l^{[\omega(.); \lambda_.(.)]}(\sigma_i^x C) }{l^{[\omega(.); \lambda_.(.)]}(C)} 
\label{wjumpforwardklargedevsingleflip}
\end{eqnarray}
while the diagonal elements are given by
 Eq. \ref{wjumpforwardklargedevdiag} 
\begin{eqnarray}
w^{Cond[\omega(.); \lambda_.(.)}(C,C) && =  \omega(C)-N- G[\omega(.); \lambda_.(.)]
\label{wjumpforwardklargedevdiagsingleflip}
\end{eqnarray}

For arbitrary functions $[\omega(.); \lambda_.(.) ]$,
one cannot solve explicitly the eigenvalue Eq. \ref{eigenleftsinglespinflip} in the space of the $2^N$ configurations,
and thus the conditioned generator of Eqs \ref{wjumpforwardklargedevsingleflip}
and \ref{wjumpforwardklargedevdiagsingleflip} is not explicit.
However there are special soluble cases, in particular in one dimension \cite{JackSollich2010},
for instance when the generator $w$ is conditioned with respect to the total Ising energy,
since the deformed generator then coincides with the quantum Hamiltonian corresponding to the Transverse-Field-Ising-Chain
(see \cite{JackSollich2010} for more details).


\subsection{ Inverse problem : functions $[\omega(C),\lambda_i(C)]$ to produce 
the given generator ${\mathring w}(.,.)=w^{Cond[\omega(.); \lambda_.(.)]}(.,.)$} 

For the inverse problem involving the two given generators of Eqs \ref{singlespinflip}
and \ref{singlespinflipindep},
the general solution of section \ref{sec_Inverse} yields:

(i) the simple solution of Eq. \ref{wCondconservedsolinverse} corresponding
to the vanishing eigenvalue $G[{\mathring \omega}(.); {\mathring \lambda}_.(.)] =0$
and 
to the trivial left eigenvector $l^{[{\mathring \omega}(.); {\mathring \lambda}_.(.)] }(x)=1 $
reduces to
\begin{eqnarray}
{\mathring \lambda}_i(C) && 
=   \ln \left( {\mathring w}_i(C) \right)
 \nonumber \\
 {\mathring \omega}(C) && = N- \sum_{i=1}^N {\mathring w}_i(C)
\label{wCondconservedsolinversesinglespin}
\end{eqnarray}

(ii) the other solutions can be parametrized by the eigenvalue $G$ that determines 
the function $\omega(C) $ via Eq. \ref{omegaomegaringsolugene}
\begin{eqnarray}
\omega^{[G]}(C)  =  G+N- \sum_{i=1}^N {\mathring w}_i(C)
\label{omegaxtsinglespin}
\end{eqnarray}
and by the positive left eigenvector $l(C)>0$ that determines the functions 
$\lambda_i(C) $ via Eq. \ref{lambdalambdaringsolugene}
\begin{eqnarray}
\lambda^{[l(.)] }_i (C) =   \ln \left( {\mathring w}_i(C) \right)  - \ln \left( l(\sigma_i^x C)  \right) + \ln \left( l(C)  \right)
 \label{lambdalambdaringsolugenesinglespin}
\end{eqnarray}

In these spatially-extended spin models, it is thus very important to distinguish two cases :

(1) if the single-spin-flip rate ${\mathring w}_i(C) $ of the spin $i$ only involves the neighboring spins of $i$
as in models with short-ranged interactions, then the simple solution of Eq \ref{wCondconservedsolinversesinglespin} will also involve only a few spins and not the whole configuration $C$. 
To be more concrete, we will consider in section \ref{sec_TwoSpinFlip} the specific case with space-local two-spin-flip dynamics in one dimension.

(2) if the single-spin-flip rate ${\mathring w}_i(C) $ of the spin $i$ involves the whole configuration $C$,
i.e. all the other spins $S_{j \ne i}$ as in models with long-ranged interactions, 
then the simple solution of Eq \ref{wCondconservedsolinversesinglespin} will also involve the whole configuration $C$. Let us now describe the simplest mean-field example.


\subsubsection*{ Simplest example when ${\mathring w} $ satisfies detailed-balance for the mean-field Curie-Weiss model at inverse-temperature $\beta$} 

In the mean-field Curie-Weiss model, the configuration $C=\{S_1,..S_{i-1},S_i,S_{i+1} ...,S_N\}$
of $N$ spins is characterized by its number $N_+(C)$ of up-spins and its number $N_-(C)$ of down-spins
that can be rewritten
\begin{eqnarray}
N_+(C) && = \sum_{i=1}^N \delta_{S_i,+} = \frac{ N+M(C)}{2}
\nonumber \\
N_-(C) && = \sum_{i=1}^N \delta_{S_i,-} = \frac{ N-M(C)}{2}
\label{CurieNpNn}
\end{eqnarray}
in terms of the magnetization
\begin{eqnarray}
M(C) = \sum_{i=1}^N S_i
\label{CurieMagn}
\end{eqnarray}
that can be used to define the ferromagnetic energy of coupling $J>0$ for the configuration $C$
\begin{eqnarray}
E(C) = - \frac{ J M^2(C)} {2N} = - \frac{ J } {2N} \sum_{i=1}^N \sum_{j=1}^N S_i S_j
\label{CurieEner}
\end{eqnarray}

In order to satisfy detailed balance via the single-spin-flip dynamics ${\mathring w} $
\begin{eqnarray}
{\mathring w} (\sigma_i^x C,C ) P_{\beta}(C) = {\mathring w} ( C,\sigma_i^xC ) P_{\beta}(\sigma_i^xC)
\label{CurieDB}
\end{eqnarray}
with respect to the Boltzmann distribution at inverse-temperature $\beta$ 
\begin{eqnarray}
P_{\beta}(C) = \frac{ e^{- \beta E(C)} }{ \displaystyle \sum_{C'} e^{- \beta E(C)}} 
\label{CurieBoltz}
\end{eqnarray}
let us consider the simple rates
\begin{eqnarray}
{\mathring w} (\sigma_i^x C,C ) &&  =  \sqrt { \frac{P_{\beta}(\sigma_i^xC)}{P_{\beta}(C)} }
= e^{ \frac{\beta}{2} \left[  E(C)- E(\sigma_i^x C) \right]} 
 = e^{  \frac{ \beta J } {N} [ 1- S_i M(C) ]}
= e^{ \displaystyle  - \frac{ \beta J } {N} S_i \sum_{j \ne i}  S_j }
\label{choiceDBCurie}
\end{eqnarray}
in order to obtain their interpretation as conditioning of the rates $w$ of Eq. \ref{singlespinflip}
that corresponds to the infinite-temperature case $\beta=0$ in Eq. \ref{choiceDBCurie}.

Plugging the first expression of Eq. \ref{choiceDBCurie} into Eq. \ref{lambdalambdaringsolugenesinglespin}
\begin{eqnarray}
\lambda^{[l(.)] }_i (C) 
&& =   \ln \left( {\mathring w}_i(C) \right)  - \ln \left( l(\sigma_i^x C)  \right) + \ln \left( l(C)  \right)
=  \ln \left(  \sqrt { \frac{P_{\beta}(\sigma_i^xC)}{P_{\beta}(C)} }  \right)  - \ln \left( l(\sigma_i^x C)  \right) + \ln \left( l(C)  \right)
\nonumber \\
&& = 0 \ \ \text{ for the choice } \ \ l(C) = \sqrt{P_{\beta}(C)}
 \label{lambdalambdaringsolugenesinglespinCurie}
\end{eqnarray}
yields that one can choose the left eigenvector $l(C) = \sqrt{P_{\beta}(C)}  $ in order to make vanish all the parameters $\lambda^{[l(.)] }_i (C)  =0 $ conjugated to the empirical flows.
The solution of Eq. \ref{omegaxtsinglespin} reads the rates of Eq. \ref{choiceDBCurie}
using Eq. \ref{CurieNpNn}
\begin{eqnarray}
\omega^{[G]}(C)  &&=  G+N- \sum_{i=1}^N {\mathring w} (\sigma_i^x C,C )
 =  G+N- \sum_{i=1}^N  e^{  \frac{ \beta J } {N} [ 1- S_i M(C) ]}
 = G+N- N_+(C) e^{  \frac{ \beta J } {N} [ 1-  M(C) ]}- N_-(C) e^{  \frac{ \beta J } {N} [ 1+  M(C) ]}
 \nonumber \\
 && = G+N- \frac{ N+M(C)}{2} e^{  \frac{ \beta J } {N} [ 1-  M(C) ]}
 - \frac{ N-M(C)}{2} e^{  \frac{ \beta J } {N} [ 1+  M(C) ]}
\label{omegaxtsinglespinCurie}
\end{eqnarray}
which is an even function of the magnetization $M(C) $,
and can thus also be considered as a function of the energy $E(C)= - \frac{ J M^2(C)} {2N}$ of Eq. \ref{CurieEner}.
In conclusion, the detailed-balance dynamics at inverse-temperature $\beta$ of Eq. \ref{choiceDBCurie}
can be considered as the conditioning of the independent-spin-flip dynamics $w$ of Eq. \ref{singlespinflip}
(corresponding to the infinite-temperature case $\beta=0$)
with respect to the empirical density $P(C)$ of configurations 
via the conjugated parameters of Eq. \ref{omegaxtsinglespinCurie} that depend only on the energy $E(C)$
of configurations.


\section{ Application to space-local two-spin-flip dynamics in classical spin chains } 

\label{sec_TwoSpinFlip}

The local two-spin-flip dynamics in a periodic chain of $N$ classical spins
can be interpreted as partially asymmetric exclusion processes with dimer evaporation and dimer deposition  \cite{Harris_twospinflip}, and also appear in the Domain-Wall formulation of the Glauber single-spin-flip dynamics of the Ising chain \cite{JackSollich2010,GlauberIsing}.

\subsection{ Notations for space-local two-spin-flip dynamics in a periodic chain of $N$ classical spins } 

For a chain of $N$ classical spins $S_j=\pm 1$ with periodic boundary conditions,
it is interesting to consider the general space-local two-spin-flip dynamics \cite{Harris_twospinflip}
that involves arbitrary rates ${\mathring w}_{i+\frac{1}{2}}^{S_i S_{i+1}}$ for the joint-flip $(S_i,S_{i+1}) \to (-S_i,-S_{i+1})$
of the two neighboring spins $S_i$ and $S_{i+1}$ independently of the values of the other spins $S_{j \ne (i,i+1)}$ of the configuration $C=(S_1,...,S_N)$ 
\begin{eqnarray}
{\mathring w} (\sigma_i^{-S_i} \sigma_{i+1}^{-S_{i+1}} C,C )   = {\mathring w}_{i+\frac{1}{2}}^{S_i S_{i+1}}
\label{twospinflipratesC}
\end{eqnarray}
 The full generator ${\mathring w}  $ then reads when written as a quantum operator using Pauli matrices 
\begin{eqnarray}
{\mathring w}   && = \sum_{i=1}^N 
\bigg[ {\mathring w}_{i+\frac{1}{2}}^{-+}\left( \sigma_i^+ \sigma_{i+1}^- -  \frac{(1-\sigma_i^z) (1+\sigma_{i+1}^z)}{4}\right)
+ {\mathring w}_{i+\frac{1}{2}}^{+-}\left( \sigma_i^- \sigma_{i+1}^+ -  \frac{(1+\sigma_i^z) (1-\sigma_{i+1}^z)}{4}\right)
\nonumber \\
&& +{\mathring w}_{i+\frac{1}{2}}^{++}\left( \sigma_i^- \sigma_{i+1}^- -  \frac{(1+\sigma_i^z) (1+\sigma_{i+1}^z)}{4}\right)
+ {\mathring w}_{i+\frac{1}{2}}^{--}\left( \sigma_i^+ \sigma_{i+1}^+ -  \frac{(1-\sigma_i^z) (1-\sigma_{i+1}^z)}{4}\right)
\bigg]
\label{twosinglespinflipPauli}
\end{eqnarray}
So the off-diagonal part ${\mathring w}^{off} $ of ${\mathring w}$ reduces to the terms involving the ladder Pauli operators $\sigma_j^{\pm}$ of two consecutive spins
\begin{eqnarray}
{\mathring w}^{off}   && = \sum_{i=1}^N 
\bigg[ {\mathring w}_{i+\frac{1}{2}}^{-+} \sigma_i^+ \sigma_{i+1}^- 
+ {\mathring w}_{i+\frac{1}{2}}^{+-} \sigma_i^- \sigma_{i+1}^+ 
 +{\mathring w}_{i+\frac{1}{2}}^{++} \sigma_i^- \sigma_{i+1}^- 
+ {\mathring w}_{i+\frac{1}{2}}^{--} \sigma_i^+ \sigma_{i+1}^+ 
\bigg]
\label{twosinglespinflipPaulioff}
\end{eqnarray}
while the diagonal part ${\mathring w}^{diag} $ involving the Pauli matrices $\sigma_j^z$ can be rewritten as
\begin{eqnarray}
{\mathring w}^{diag}   && =  - {\mathring K} + \sum_{i=1}^N \left[ {\mathring h}_i \sigma_i^z + {\mathring J}_{i+\frac{1}{2}} \sigma_i^z\sigma_{i+1}^z\right]
\label{twosinglespinflipPaulidiag}
\end{eqnarray}
where we have introduced the following notations for
 the constant
\begin{eqnarray}
{\mathring K}   && \equiv  \sum_{i=1}^N 
 \frac{{\mathring w}_{i+\frac{1}{2}}^{-+}  + {\mathring w}_{i+\frac{1}{2}}^{+-} 
 +{\mathring w}_{i+\frac{1}{2}}^{++} + {\mathring w}_{i+\frac{1}{2}}^{--}   }{4}
\label{twosinglespinflipPaulidiagcte}
\end{eqnarray}
for the magnetic fields
\begin{eqnarray}
{\mathring h}_i   \equiv 
 \frac{ ({\mathring w}_{i+\frac{1}{2}}^{-+} - {\mathring w}_{i-\frac{1}{2}}^{-+}) 
 + ( {\mathring w}_{i-\frac{1}{2}}^{+-} - {\mathring w}_{i+\frac{1}{2}}^{+-} )
 - ( {\mathring w}_{i-\frac{1}{2}}^{++}+ {\mathring w}_{i+\frac{1}{2}}^{++}) + ({\mathring w}_{i-\frac{1}{2}}^{--} + {\mathring w}_{i+\frac{1}{2}}^{--})   }{4}
\label{twosinglespinflipPaulidiaghz}
\end{eqnarray}
and for the couplings
\begin{eqnarray}
{\mathring J}_{i+\frac{1}{2}} \equiv \frac{{\mathring w}_{i+\frac{1}{2}}^{-+}  + {\mathring w}_{i+\frac{1}{2}}^{+-} 
 - {\mathring w}_{i+\frac{1}{2}}^{++} - {\mathring w}_{i+\frac{1}{2}}^{--}   }{4}
\label{twosinglespinflipPaulidiagJzz}
\end{eqnarray}
Note that when all these couplings vanish ${\mathring J}_{i+\frac{1}{2}} =0 $,
then the generator ${\mathring w} $
corresponds to a quantum Hamiltonian that can be diagonalized via free-fermions \cite{Harris_twospinflip}.

The simplest dynamics with the same possible transitions as in Eq. \ref{twospinflipratesC}
corresponds to the generator $w $
where all the two-spin-flip rates are instead unity
\begin{eqnarray}
w (\sigma_i^{-S_i} \sigma_{i+1}^{-S_{i+1}} C,C )  && = w_{i+\frac{1}{2}}^{S_i S_{i+1}}  =1
\nonumber \\
w ( C,C )  && = -N
\label{twosinglespinflipratesunity}
\end{eqnarray}
so that the total generator $w$ when written as a quantum operator using Pauli matrices as in Eq. \ref{twosinglespinflipPauli}
\begin{eqnarray}
w  = -N + \sum_{i=1}^N 
\bigg[  \sigma_i^+ \sigma_{i+1}^- 
+  \sigma_i^- \sigma_{i+1}^+ 
 + \sigma_i^- \sigma_{i+1}^- 
+  \sigma_i^+ \sigma_{i+1}^+ 
\bigg]
 = \sum_{i=1}^N \bigg[  \sigma_i^x \sigma_{i+1}^x  -1 \bigg]
\label{wtwosinglespinflipPauli}
\end{eqnarray}
corresponds to the independent flip of pairs of neighboring spins with rate unity.
The generator $w$ satisfies detailed-balance with respect to its uniform steady state 
$P_*(C)=\frac{1}{2^N}$ over the $2^N$ configurations.


\subsection{ Inverse problem : functions $[\omega(C),\lambda(\sigma_i^{-S_i} \sigma_{i+1}^{-S_{i+1}} C, C)]$ to produce 
the given generator ${\mathring w}(.,.)=w^{Cond[\omega(.); \lambda(.,.)]}(.,.)$}

For the inverse problem concerning the two generators of Eqs \ref{twospinflipratesC}
and \ref{twosinglespinflipratesunity},
the simple solution of Eq. \ref{wCondconservedsolinverse} for the configuration $C=(S_1,...,S_N)$
\begin{eqnarray}
{\mathring \lambda}(\sigma_i^{-S_i} \sigma_{i+1}^{-S_{i+1}} C,C) 
&& = \ln \left( {\mathring w}_{i+\frac{1}{2}}^{S_i S_{i+1}} \right)
 \nonumber \\
 {\mathring \omega}(C) && = N   - {\mathring K} + \sum_{i=1}^N \left[ {\mathring h}_i S_i + {\mathring J}_{i+\frac{1}{2}} S_i S_{i+1}\right]
\label{wCondconservedsolinversetwospin}
\end{eqnarray}
only involves local information on the configuration $C=(S_1,...,S_N)$.
It is thus interesting to rewrite the conditioning factors that are in the exponential of the generating function of Eq. \ref{geneddef}
 as follows :

(1) the first factor involving the empirical probability $P(S_1,..,S_N)$ of Eq. \ref{rho1pj} 
for the configuration $C= \{S_1,..,S_N\}$
\begin{eqnarray}
  P(S_1,..,S_N)  \equiv \frac{1}{T} \int_0^T dt \  \delta_{S_1(t),S_1}\delta_{S_2(t),S_2}   ... \delta_{S_N(t),S_N} 
 \label{rhospin}
\end{eqnarray}
satisfying the normalization of Eq. \ref{rho1ptnormaj}
\begin{eqnarray}
  \sum_{S_1,..,S_N  }  P(S_1,..,S_N) =1
\label{rhospinnorma}
\end{eqnarray}
and the conjugated variables ${\mathring \omega}( C) $ of Eq. \ref{wCondconservedsolinversetwospin}
can be rewritten as
\begin{eqnarray}
 \sum_C P(C)  {\mathring \omega}(C) && = \sum_{S_1,..,S_N  }  
 P(S_1,..,S_N) \left( N   - {\mathring K} + \sum_{i=1}^N \left[ {\mathring h}_i S_i + {\mathring J}_{i+\frac{1}{2}} S_i S_{i+1}\right]
 \right)
\nonumber \\
&& =  N   - {\mathring K} + \sum_{i=1}^N  {\mathring h}_i \sum_{S_i=\pm} S_i P_i^{S_i} 
+ \sum_{i=1}^N {\mathring J}_{i+\frac{1}{2}} \sum_{S_i=\pm} \sum_{S_{i+1}=\pm} S_i S_{i+1} P_{i,i+1}^{S_i S_{i+1}}  
\label{trajobs2spinempi1}
\end{eqnarray}
that only involves the empirical densities concerning a single spin
\begin{eqnarray}
P_i^{S_i}   \equiv \frac{1}{T} \int_0^T dt \  \delta_{S_i(t),S_i}
 \label{rhospin1}
\end{eqnarray}
or two consecutive spins
\begin{eqnarray}
P_{i,i+1}^{S_i S_{i+1}}  \equiv \frac{1}{T} \int_0^T dt \  \delta_{S_i(t),S_i}\delta_{S_{i+1}(t),S_{i+1}}  
 \label{rhospin2}
\end{eqnarray}
and not the empirical density $P(C)=P(S_1,..,S_N)$ of the whole configuration anymore.

(2) the second factor involving the two-spin-flip flows of Eq. \ref{jumpempiricaldensity} for the configurations
$C=(S_1,...,S_N)$ 
\begin{eqnarray}
Q(\sigma_i^{-S_i} \sigma_{i+1}^{-S_{i+1}} C,C)  \equiv  \frac{1}{T} 
\sum_{\substack{ t \in [0,T] : S_i(t^+)=-S_i \text{ and }  S_i(t^-)=S_i \\
S_{i+1}(t^+)=-S_{i+1} \text{ and } S_{i+1}(t^-)=S_{i+1}}} \prod_{j \ne (i,i+1)} \delta_{S_j(t),S_j} 
\label{jumpempiricaldensityspin}
\end{eqnarray}
and the conjugated variables ${\mathring \lambda}(\sigma_i^{-S_i} \sigma_{i+1}^{-S_{i+1}} C,C)  $ of Eq. \ref{wCondconservedsolinversetwospin}
can be rewritten as
\begin{eqnarray}
&& \sum_{S_1,..,S_N  }   \sum_{i=1}^N Q(\sigma_i^{-S_i} \sigma_{i+1}^{-S_{i+1}} C,C) 
   {\mathring \lambda}(\sigma_i^{-S_i} \sigma_{i+1}^{-S_{i+1}} C,C) 
    =   \sum_{S_1,..,S_N  }  \sum_{i=1}^N Q(\sigma_i^{-S_i} \sigma_{i+1}^{-S_{i+1}} C,C) 
  \ln \left( {\mathring w}_{i+\frac{1}{2}}^{S_i S_{i+1}} \right)
\nonumber \\
&& = \sum_{i=1}^N \sum_{S_i=\pm} \sum_{S_{i+1}=\pm} Q_{i,i+1}^{S_i,S_{i+1}} 
  \ln \left( {\mathring w}_{i+\frac{1}{2}}^{S_i S_{i+1}} \right)
\label{trajobs2spinempi2}
\end{eqnarray}
that only involves the empirical two-spin-flip of two consecutive spins
\begin{eqnarray}
Q_{i,i+1}^{S_i,S_{i+1}}  \equiv  \frac{1}{T} 
\sum_{\substack{ t \in [0,T] : S_i(t^+)=-S_i \text{ and }  S_i(t^-)=S_i \\
S_{i+1}(t^+)=-S_{i+1} \text{ and } S_{i+1}(t^-)=S_{i+1}}}  1
\label{jumpempiricaldensityspin2}
\end{eqnarray}
independently of the values of the other spins $S_{j \ne i,i+1}$ of the configuration that were in the empirical flows of Eq. \ref{jumpempiricaldensityspin}.

Putting Eqs \ref{trajobs2spinempi1} and \ref{trajobs2spinempi2} together, one obtains that 
the conditioning observable
associated to the simple solution of Eq. \ref{wCondconservedsolinversetwospin}
only involves empirical observables concerning a single spin or two consecutive spins
as given by Eqs \ref{rhospin1}
\ref{rhospin2}
and \ref{jumpempiricaldensityspin2}
\begin{eqnarray}
&&         \sum_{S_1,..,S_N  }  \left[  P(S_1,..,S_N) {\mathring \omega}(S_1,..,S_N)
        +  \sum_{i=1}^N Q(\sigma_i^{-S_i} \sigma_{i+1}^{-S_{i+1}} C,C) 
   {\mathring \lambda}(\sigma_i^{-S_i} \sigma_{i+1}^{-S_{i+1}} C,C)
\right]
\nonumber \\
&& = N   - {\mathring K} + \sum_{i=1}^N  {\mathring h}_i  \sum_{S_i=\pm} S_i P_i^{S_i} 
+ \sum_{i=1}^N {\mathring J}_{i+\frac{1}{2}} \sum_{S_i=\pm} \sum_{S_{i+1}=\pm} S_i S_{i+1} P_{i,i+1}^{S_i S_{i+1}} 
        + \sum_{i=1}^N \sum_{S_i=\pm} \sum_{S_{i+1}=\pm} Q_{i,i+1}^{S_i,S_{i+1}} 
  \ln \left( {\mathring w}_{i+\frac{1}{2}}^{S_i S_{i+1}} \right)
\label{geneddeftwospin}
\end{eqnarray}
To be more concrete, let us now focus on the case with translation-invariance.


\subsubsection{ Special case with translation-invariance of the rates 
${\mathring w}_{i+\frac{1}{2}}^{S S'}={\mathring w}^{S S'}$ along the ring } 

When the kernel ${\mathring w} $ of Eqs \ref{twosinglespinflipPaulioff}
and \ref{twosinglespinflipPaulidiag}
is translation-invariant and involves only the four parameters ${\mathring w}^{S=\pm,S'=\pm} $
\begin{eqnarray}
{\mathring w}^{off}   && = \sum_{i=1}^N 
\bigg[ {\mathring w}^{-+} \sigma_i^+ \sigma_{i+1}^- 
+ {\mathring w}^{+-} \sigma_i^- \sigma_{i+1}^+ 
 +{\mathring w}^{++} \sigma_i^- \sigma_{i+1}^- 
+ {\mathring w}^{--} \sigma_i^+ \sigma_{i+1}^+ 
\bigg]
\nonumber \\
{\mathring w}^{diag}   && =  - {\mathring K} 
+ \sum_{i=1}^N \left[ {\mathring h} \sigma_i^z + {\mathring J} \sigma_i^z\sigma_{i+1}^z\right]
\label{twosinglespinflipPaulidiagtranslation}
\end{eqnarray}
where the notations of Eqs \ref{twosinglespinflipPaulidiagcte}
\ref{twosinglespinflipPaulidiaghz}
\ref{twosinglespinflipPaulidiagJzz} simplify into
\begin{eqnarray}
{\mathring K}   && \equiv  N
 \frac{{\mathring w}^{-+}  + {\mathring w}^{+-} 
 +{\mathring w}^{++} + {\mathring w}^{--}   }{4}
\nonumber \\
{\mathring h} &&  \equiv 
 \frac{ {\mathring w}^{--} - {\mathring w}^{++}  }{2}
\nonumber \\
{\mathring J} && \equiv \frac{{\mathring w}^{-+}  + {\mathring w}^{+-} 
 - {\mathring w}^{++} - {\mathring w}^{--}   }{4}
\label{twosinglespinflipPaulidiagJzztranslation}
\end{eqnarray}
the conditioning observable of Eq. \ref{geneddeftwospin} 
\begin{eqnarray}
  N   - {\mathring K} +  {\mathring h} M^{tot}
+{\mathring J} C^{tot}_{nn}
        + Q^{++}  \ln \left( {\mathring w}^{++} \right) 
        +  Q^{--}  \ln \left( {\mathring w}^{--} \right)
        + Q^{+-}  \ln \left( {\mathring w}^{+-} \right) 
        +  Q^{-+}  \ln \left( {\mathring w}^{-+} \right)   
\label{geneddeftwospintranslation}
\end{eqnarray}
involves the total empirical magnetization  
\begin{eqnarray}
M^{tot} \equiv \sum_{i=1}^N   \sum_{S_i=\pm} S_i P_i^{S_i}  =\sum_{i=1}^N  (P_i^{+} -P_i^-)
\label{empiricalMtranslation}
\end{eqnarray}
the total empirical correlation between nearest-neighboring sites $(i,i+1)$
\begin{eqnarray}
C^{tot}_{nn} \equiv  \sum_{i=1}^N \sum_{S_i=\pm} \sum_{S_{i+1}=\pm} S_i S_{i+1} P_{i,i+1}^{S_i S_{i+1}} = \sum_{i=1}^N  (P_{i,i+1}^{++} +P_{i,i+1}^{--} - P_{i,i+1}^{+-} - P_{i,i+1}^{-+})
\label{empiricalCnntranslation}
\end{eqnarray}
and the four types of global empirical flows $Q^{S=\pm 1,S'=\pm 1} $ on the ring
\begin{eqnarray}
Q^{S,S'}   \equiv \sum_{i=1}^N  Q_{i,i+1}^{S,S'}  
\label{empiricalQtranslation}
\end{eqnarray}

In the two next subsections, we describe two special choices of the four parameters ${\mathring w}^{S=\pm,S'=\pm} $
that will correspond for ${\mathring w} $ to an equilibrium dynamics or to an out-of-equilibrium dynamics respectively.


\subsubsection{ Example of the Domain-Wall formulation of the detailed-balance Glauber dynamics of the pure Ising chain } 

\label{subsec_glauber}

The Domain-Wall formulation of the detailed-balance Glauber single-spin-flip dynamics of the pure Ising chain at inverse temperature $\beta$ corresponds to the parameters \cite{JackSollich2010,GlauberIsing}
\begin{eqnarray}
 {\mathring w}^{-+} && = 1
 \nonumber \\
  {\mathring w}^{+-} && = 1
 \nonumber \\
 {\mathring w}^{++}  && = \frac{2 e^{-\beta}}{ e^{\beta} +e^{-\beta} }
 \nonumber \\
 {\mathring w}^{--} && = \frac{2 e^{\beta}}{ e^{\beta} +e^{-\beta} }
 \label{glauber}
\end{eqnarray}
that can be plugged into Eq. \ref{twosinglespinflipPaulidiagJzztranslation}
\begin{eqnarray}
{\mathring K}   && =  N
\nonumber \\
{\mathring h} &&  = \frac{ e^{\beta}- e^{-\beta}}{ e^{\beta} +e^{-\beta} } = \tanh (\beta) 
\nonumber \\
{\mathring J} && =0
\label{twosinglespinflipPaulidiagJzztranslationGlauber}
\end{eqnarray}
to obtain that the conditioning observable of Eq. \ref{geneddeftwospintranslation}
\begin{eqnarray}
   M^{tot}  \tanh (\beta) 
        + Q^{++}  \ln \left(  \frac{2 e^{-\beta}}{ e^{\beta} +e^{-\beta} } \right) 
        +  Q^{--}  \ln \left( \frac{2 e^{\beta}}{ e^{\beta} +e^{-\beta} } \right)           
\label{geneddeftwospintranslationGlauber}
\end{eqnarray}
only involves the empirical magnetization $ M^{tot}$,
 the empirical flow $Q^{++} $ associated to the creation of two neighboring domain-walls,
and the empirical flow $Q^{--} $ associated to the annihilation of two neighboring domain-walls,
with conjugated parameters that depend on the inverse-temperature $\beta$.
Note that in the present section, we have chosen to focus only the simplest solution 
$[{\mathring \omega}(.); {\mathring \lambda}(.,.)] $,
so if one wishes to obtain all the other conditionings
able to produce the equilibrium Glauber dynamics, one can analyze
the corresponding general solution of the inverse problem.


\subsubsection{ Example of the pure non-equilibrium partially asymmetric exclusion processes with dimer evaporation and deposition} 

\label{subsec_exclusion}

The pure non-equilibrium partially asymmetric exclusion processes with dimer evaporation and dimer deposition  \cite{Harris_twospinflip} for the following values of the four parameters in terms of the single parameter $q \in ]1,2[ $
\begin{eqnarray}
 {\mathring w}^{-+} && = q \in ]1,2[
 \nonumber \\
  {\mathring w}^{+-} && = (2-q) \in ]0,1[
 \nonumber \\
 {\mathring w}^{++}  && = 1
 \nonumber \\
 {\mathring w}^{--} && = 1
 \label{exclusion}
\end{eqnarray}
correspond to the following values in Eq. \ref{twosinglespinflipPaulidiagJzztranslation}
\begin{eqnarray}
{\mathring K}   && =  N 
\nonumber \\
{\mathring h} &&  =0
\nonumber \\
{\mathring J} && =0
\label{twosinglespinflipPaulidiagJzztranslationexclusion}
\end{eqnarray}
As a consequence, the conditioning observable of Eq. \ref{geneddeftwospintranslation}
only involves the empirical flow $Q^{-+} $ counting the total number of hoppings in the right direction
and the empirical flow $Q^{+-} $ counting the total number of hoppings in the left direction
\begin{eqnarray}
 Q^{-+}  \ln \left( q \right) + Q^{+-}  \ln \left( 2-q \right)
\label{geneddeftwospintranslationexclusion}
\end{eqnarray}
The rewriting in terms of the empirical current ${\cal J}=Q^{-+} - Q^{+-}$ flowing around the ring
and in terms of the total empirical number ${\cal A}=Q^{-+} + Q^{+-}$ of hoppings
\begin{eqnarray}
 Q^{-+}  && = \frac{{\cal A} + {\cal J}}{2}
 \nonumber \\
  Q^{+-}  && = \frac{{\cal A} - {\cal J}}{2}
\label{exclusionQJN}
\end{eqnarray}
yields that the conditioning observable of Eq. \ref{geneddeftwospintranslationexclusion} 
\begin{eqnarray}
 {\cal A}  \frac{ \ln \left( q ( 2-q ) \right) }{2}   
         +   {\cal J}  \frac{ \ln \left( \frac{q }{ 2-q } \right)  }{2} 
\label{geneddeftwospintranslationexclusionaj}
\end{eqnarray}
involves the total empirical number ${\cal A}=Q^{-+} + Q^{+-}$ of hoppings
and to the empirical current ${\cal J}=Q^{-+} - Q^{+-}$ flowing around the ring
with conjugated parameters given in terms of the asymmetry parameter $q \in ]0,1[$ of the
non-equilibrium partially asymmetric exclusion process of Eq. \ref{exclusion}.


\section{ Conclusions } 

\label{sec_conclusion}

After recalling the direct problem of the canonical conditioning of a given Markov process with respect to a given time-local trajectory observable in relation with the spontaneous fluctuations of the empirical density and of the empirical flows existing in the initial dynamics, we have analyzed the following inverse problem: when two Markov generators are given, is it possible to connect them via some canonical conditioning and to construct the corresponding time-local trajectory observable?
For clarity, we have focused on continuous-time Markov jump processes in discrete configuration space in the main text and on diffusions processes in continuous space in the Appendices. 

We have first stressed the following necessary and sufficient conditions for the inverse problem : 

(i) for continuous-time Markov jump processes, the two generators should involve the same possible elementary jumps in configuration space, i.e. only the values of the corresponding rates can differ; 

(ii) for diffusion processes, the two Fokker-Planck generators should involve the same diffusion coefficient, i.e. only the two forces can differ.

In both settings, we have then constructed explicitly the various time-local trajectory observables that can be used to connect the two given generators via canonical conditioning. Finally, we have illustrated this general framework with various applications involving a single particle or many-body spin models. 
In particular, in order to contribute to the much debated question of whether non-equilibrium dynamics can be considered as some conditionings of equilibrium dynamics \cite{Evans2004,Evans2005,Evans2008,Baule2008,Evans2010,JackSollich2010,c_maximizationDynEntropy,Andrieux2012,JackEvans,chetrite_optimal,verley2016,MaximumCaliber,verley2022,Andrieux2022},
we have described several examples where non-equilibrium Markov processes with non-vanishing steady currents can be interpreted as the canonical conditionings of detailed-balance processes with respect to explicit time-local trajectory observables (see sections \ref{sec_UnDirected} and \ref{subsec_exclusion}
as well as Appendices \ref{app_Ring} and \ref{app_Diffd}).
We have also described other examples 
where both generators correspond to non-equilibrium dynamics (see section \ref{sec_Directed})
and where both generators correspond to equilibrium dynamics (see the final example in section \ref{sec_SingleSpinFlip} and subsection \ref{subsec_glauber}).
 In most of these concrete applications, 
 we have explained in detail how one can use the freedom in the choice of the solution
 to the inverse problem for the two functions $ [\omega(.); \lambda(.,.)]$
 in order to have a simpler physical meaning for the corresponding conditioning observable.


\appendix

\section{ Reminder on canonical conditioning for Diffusion processes  in dimension $d$}

\label{app_Diffusion}

In this Appendix, we recall the direct problem of the canonical condition of a given diffusion process
with respect to trajectory observables in relation with the dynamical large deviations of the empirical density and the empirical current. The ideas are the same as in the section \ref{sec_jump} of the main text concerning Markov jump processes,
but there are some important technical differences that need to be emphasized.

\subsection{ Fokker-Planck generator ${\cal F}$ involving the forces $F_{\mu}(\vec x) $ and the diffusion coefficients $D_{\mu }(\vec x)$} 

The Fokker-Planck dynamics in dimension $d$
for the probability density $\rho^{[t]}(\vec x)$
to be at position $\vec x $ at time $t$
\begin{eqnarray}
 \partial_t \rho^{[t]}(\vec x)  = {\cal F} \rho^{[t]}(\vec x)
\label{fokkerplanck}
\end{eqnarray}
involves the Fokker-Planck generator
\begin{eqnarray}
 {\cal F}  \equiv  -   \sum_{\mu=1}^d \frac{ \partial  }{\partial x_{\mu} } \left[  F_{\mu} (\vec x )  -D_{\mu} (\vec x)  \frac{\partial }{ \partial x_{\mu}}   
\right]
\label{fokkerplanckgenerator}
\end{eqnarray}
So Eq. \ref{fokkerplanck}
corresponds to a continuity equation for the probability density $\rho^{[t]}(\vec x)$
\begin{eqnarray}
 \partial_t \rho^{[t]}(\vec x)  =  -   \vec \nabla . \vec j^{[t]} (\vec x)
 \equiv  -   \sum_{\mu=1}^d \frac{ \partial  j^{[t]}_{\mu} (\vec x) }{\partial x_{\mu} }
\label{fokkerplanckcontinuity}
\end{eqnarray}
where the current component $j^{[t]}_{\mu} (\vec x) $ in the direction $\mu \in \{1,..,d\}$
involves the force component $F_{\mu}(\vec x)  $ and
the diffusion coefficient $D_{\mu}(\vec x) $ 
\begin{eqnarray}
 j^{[t]}_{\mu} (\vec x) \equiv  F_{\mu} (\vec x ) \rho^{[t]}(\vec x) -D_{\mu} (\vec x)  \frac{\partial \rho^{[t]}(\vec x)}{ \partial x_{\mu}}   
\label{fokkerplanckj}
\end{eqnarray}

We assume that this dynamics converges towards a normalizable steady-state $\rho^*(\vec x)$ 
with its associated steady current 
\begin{eqnarray}
 j^*_{\mu} (\vec x) \equiv  F_{\mu} (\vec x ) \rho^*(\vec x) -D_{\mu} (\vec x)  \frac{\partial \rho^*(\vec x)}{ \partial x_{\mu}}  
  \label{fokkerplanckjsteady}
\end{eqnarray}
satisfying Eq. \ref{fokkerplanck}
\begin{eqnarray}
0= {\cal F} \rho^*(\vec x)  = -     \vec \nabla . \vec j^* (\vec x) \equiv 
 -   \sum_{\mu=1}^d \frac{ \partial  j^*_{\mu} (\vec x) }{\partial x_{\mu} }
\label{fokkerplancksteady}
\end{eqnarray}

From the point of view of the spectral properties of the Fokker-Planck generator ${\cal F}$,
this means that the highest eigenvalue vanishes
where the positive right eigenvector is given by the steady state 
\begin{eqnarray}
 r(x)=\rho_*(x) 
\label{FPright}
\end{eqnarray}
while the positive left eigenvector satisfying ${\cal F}^{\dagger} l(x)=0 $ is trivial
\begin{eqnarray}
 l(x)=1
\label{FPleft}
\end{eqnarray}
and associated to the conservation of probability.


\subsection{ Fokker-Planck generator as an euclidean non-hermitian electromagnetic quantum Hamiltonian $H$   }

The Fokker-Planck generator ${\cal F} $ of Eq. \ref{fokkerplanckgenerator}
can be written as the opposite of the non-hermitian quantum Hamiltonian
\begin{eqnarray}
H  = -{\cal F }   
&& = \sum_{\mu=1}^d \bigg( -  \frac{\partial }{ \partial x_{\mu}}  D_{\mu} (\vec x)  \frac{\partial }{ \partial x_{\mu}} +  F_{\mu} ( \vec x) \partial_{\mu} 
+  \frac{\partial F_{\mu} ( \vec x)}{ \partial x_{\mu}}    \bigg)
\nonumber \\
&&  =  -  \sum_{\mu=1}^d  \bigg(  \frac{\partial }{ \partial x_{\mu}}  -  A_{\mu} ( \vec x ) \bigg) D_{\mu} (\vec x) \bigg(  \frac{\partial }{ \partial x_{\mu}}  -  A_{\mu} ( \vec x ) \bigg)
    +  V(\vec x)
\label{Hamiltonian}
\end{eqnarray}
that involves the vector potential
\begin{eqnarray}
A_{\mu} ( \vec x )  \equiv \frac{  F_{\mu} ( \vec x )}{2 D_{\mu} ( \vec x )}
\label{vectorpot}
\end{eqnarray}
and the scalar potential  
\begin{eqnarray}
V(\vec x) \equiv  \sum_{\mu=1}^d \left[  D_{\mu} (\vec x) A^2_{\mu} ( \vec x ) 
+  \frac{ \partial \big(D_{\mu} (\vec x) A_{\mu} ( \vec x ) \big)}{\partial x_{\mu} }  \right]
 = \sum_{\mu=1}^d \left[ \frac{ F_{\mu}^2 (\vec x) }{ 4 D_{\mu} (\vec x)} +
 \frac{1}{2}  \frac{ \partial F_{\mu} (\vec x)}{\partial x_{\mu} } \right] 
\label{scalarpot}
\end{eqnarray}

This quantum mechanical interpretation and its consequences are discussed in detail
for the special case of constant uniform diffusion coefficient $D_{\mu}(\vec x) = \frac{1}{2} $ in \cite{us_gyrator} 
and for the present case of space-dependent diffusion coefficients $D_{\mu} (\vec x) $ 
in each direction $\mu=1,..,d$ in \cite{c_diffReg}
(where the more general case of diffusion matrix $D_{\mu \nu}(\vec x)$ can also be found in Appendices).
 Let us stress that in the present paper,
 we will only use the link between the Markov generator and the quantum Hamiltonian
of Eq. \ref{Hamiltonian}, while detailed discussions of trajectory probabilities and path-integrals involving the corresponding
classical Lagrangians can be found in \cite{us_gyrator,c_diffReg}.


\subsection{ Spontaneous fluctuations of time-averaged observables over a large-time window $[0,T]$  }

\subsubsection{ Empirical density $\rho(\vec x)$ and empirical current $\vec j(\vec x)$ 
with their constitutive constraints} 

For a diffusive trajectory $\vec x (0 \leq t \leq T) $ over a large-time-window $[0,T]$, 
 the two relevant empirical time-averaged observables are :

(i) the empirical density $\rho (.) $ that measures the fraction of the time spent at the various positions $\vec x$
\begin{eqnarray}
\rho (\vec x)  \equiv \frac{1}{T} \int_0^T dt \  \delta^{(d)} ( \vec x(t)- \vec x)  
\label{rhodiff}
\end{eqnarray}
with the normalization to unity
\begin{eqnarray}
\int d^d \vec x \ \rho (\vec x)  = 1
\label{rho1ptnormadiff}
\end{eqnarray}

(ii) the empirical current $\vec j (\vec x)$ that measures the time-averages 
of the velocity $\frac{d   \vec x (t)}{dt} $
seen at the various positions $\vec x$ within in the Stratonovich mid-point interpretation
\begin{eqnarray} 
\vec j(\vec x) \equiv   \frac{1}{T} \int_0^T dt \ \frac{d   \vec x (t)}{dt}   \delta^{(d)}( \vec x(t)- \vec x)  
\label{diffj}
\end{eqnarray}
satisfying the vanishing-divergence constraint at any position $\vec x$
\begin{eqnarray}
0 =\vec \nabla . \vec j (\vec x) = \sum_{\mu=1}^d \frac{ \partial  j_{\mu} (\vec x) }{\partial x_{\mu} }   
 \label{divergencenulle}
\end{eqnarray}


\subsubsection{ Explicit joint distribution of the empirical density $\rho(.)$ and of the empirical current $\vec j(.)$ for large $T$} 

For large $T$, the joint probability distribution 
of the normalized empirical density $\rho(.)$ of Eq. \ref{rhodiff}
and of the divergence-less empirical current $\vec j (.)$ of Eq. \ref{diffj}
satisfies the large deviation form 
\cite{wynants_thesis,maes_diffusion,chetrite_formal,engel,chetrite_HDR,c_lyapunov,c_inference,c_susyboundarydriven,c_diffReg}
\begin{eqnarray}
 p^{[2.5]}_T[ \rho(.),  \vec j (.)]   \opsimeq_{T \to +\infty}   C_{2.5} [ \rho(.),  \vec j  (.)]
e^{- \displaystyle T I_{2.5} [ \rho(.),  \vec j  (.)]    }
\label{ld2.5diff}
\end{eqnarray}
where $ C_{2.5} [ \rho(.),  \vec j  (.)] $ summarizes the constitutive constraints of Eqs \ref{rho1ptnormadiff} and \ref{divergencenulle}
\begin{eqnarray}
  C_{2.5} [ \rho(.), \vec j  (.)] =\delta \left(\int d^d \vec x \rho(\vec x) -1  \right)
\left[ \prod_{\vec x }  \delta \left(  \vec \nabla . \vec j (\vec x)    \right) \right] 
  \label{c2.25diff}
\end{eqnarray}
while the rate function at level 2.5 reads
\begin{eqnarray}
  I_{2.5}  [ \rho(.),  \vec j  (.)]  =
  \int d^d \vec x \sum_{\mu=1}^d \frac{\left[ j_{\mu}(\vec x) -   F_{\mu}(\vec x)  \rho(\vec x)
  +D_{\mu} (\vec x) \frac{\partial  \rho(\vec x)}{\partial x_{\mu} }  \right]^2}{ 4 D_{\mu} (\vec x)  \rho(\vec x) } 
  \label{rate2.25diff}
\end{eqnarray}


\subsubsection{ Different perspective when the empirical current $\vec j(.)$ is replaced by the empirical force $\vec F^E(.)$  }

A useful different perspective can be obtained via the replacement
of the empirical current $\vec j(.)$ by the empirical force $\vec F^E(.) $ 
that would have the empirical variables $[\rho(.);\vec j(.)]$ as steady state and steady current
using Eq. \ref{fokkerplanckjsteady}
\begin{eqnarray}
 j_{\mu} (\vec x) \equiv  F^E_{\mu} (\vec x ) \rho(\vec x) -D_{\mu} (\vec x)  \frac{\partial \rho(\vec x)}{ \partial x_{\mu}}  
  \label{empiricalforce}
\end{eqnarray}

Then the joint distribution of Eq. \ref{ld2.5diff} translates into the joint distribution of the empirical density $\rho(.)$
and of the empirical force $ \vec F^E (.) $
\begin{eqnarray}
 p^{[2.5]}_T[ \rho(.),  \vec F^E (.)]   \opsimeq_{T \to +\infty}   C_{2.5} [ \rho(.),  \vec F^E  (.)]
e^{- \displaystyle T I_{2.5} [ \rho(.),  \vec F^E  (.)]    }
\label{ld2.5diffinter}
\end{eqnarray}
where the rate function translated from Eq. \ref{rate2.25diff} reduces to
\begin{eqnarray}
  I_{2.5}  [ \rho(.),  \vec F^E  (.)]  =
  \int d^d \vec x \rho(\vec x) \sum_{\mu=1}^d \frac{\left[ F^E_{\mu} (\vec x )   -   F_{\mu}(\vec x) 
    \right]^2}{ 4 D_{\mu} (\vec x)   } 
  \label{rate2.25diffinfer}
\end{eqnarray}
while the constraints $ C_{2.5} [ \rho(.),  \vec F  (.)] $ translates from Eq. \ref{c2.25diff}
\begin{eqnarray}
  C_{2.5} [ \rho(.), \vec F^E  (.)] =\delta \left(\int d^d \vec x \rho(\vec x) -1  \right)
\left[ \prod_{\vec x }  \delta \left(  
    \sum_{\mu=1}^d \frac{ \partial  }{\partial x_{\mu} } \left[  F^E_{\mu} (\vec x ) \rho(\vec x) -D_{\mu} (\vec x)  \frac{\partial \rho(\vec x)}{ \partial x_{\mu}}  \right] 
    \right) \right] 
  \label{c2.25diffinfer}
\end{eqnarray}
imposes that the normalized empirical density $ \rho(.)$ should be the steady state of the empirical Fokker-Planck ${\cal F}^E $
generator associated to the empirical force $\vec F^E(.)$.
\begin{eqnarray}
{\cal F}^E \equiv -  
    \sum_{\mu=1}^d \frac{ \partial  }{\partial x_{\mu} } \left[  F^E_{\mu} (\vec x )  -D_{\mu} (\vec x)  \frac{\partial }{ \partial x_{\mu}}  \right] 
  \label{empiricalFPgenerator}
\end{eqnarray}

So Eq. \ref{ld2.5diffinter} describes how the empirical density $\rho(.)$ and the empirical Fokker-Planck
generator ${\cal F}^E $ 
as parametrized by the empirical force $\vec F^E_{\mu} (. ) $ in Eq. \ref{empiricalFPgenerator}
can fluctuate around the steady state $\rho^*(.)$ and
the true Fokker-Planck
generator ${\cal F} $ involving the true force $\vec F(.)$.


\subsection{ Generating function of the empirical density $\rho(.)$ and the empirical current $\vec j(.)$ } 

Instead of characterizing the statistics 
of the empirical density $\rho(.)$ and the empirical current $\vec j(.)$
via their joint distribution $ p^{[2.5]}_T[ \rho(.),  \vec j (.)]  $ of Eq. \ref{ld2.5diff},
one can consider their joint generating function $Z_T^{[v(.);\vec a(.)]}(\vec x \vert \vec x_0) $ 
over the Markov trajectories $\vec x(t_0 \leq s \leq t) $ starting at $\vec x(t_0)=\vec x_0$ and ending at $\vec x(t)=\vec x$
where the function $v(.) $ is conjugated to the empirical density $\rho(.)$,
while the function $ \vec a(.)$ is conjugated to the empirical current $\vec j(.)$
\begin{eqnarray}
Z_T^{[v(.);\vec a(.)]}(\vec x \vert \vec x_0) 
&& \equiv \langle \delta_{\vec x(T),\vec x} \ e^{ \displaystyle T \int d^d \vec x \left(  -v(\vec x)  \rho(\vec x)   + \vec a (\vec x). \vec j(\vec x) \right)}  \ \delta_{\vec x(0),\vec x_0} \rangle
\nonumber \\
&& =  
 \langle \delta_{\vec x(T),\vec x} \ e^{ \displaystyle \int_0^T dt \left[   -v(\vec x(t))  + \vec a (\vec x(t)).  \frac{d  \vec x(t)}{dt} \right] }  \ \delta_{\vec x(0),\vec x_0} \rangle
\label{genediffdef}
\end{eqnarray}



\subsubsection{ Consequences of the constraints on the empirical observables $[\rho(.);\vec j(.)]$ for their conjugated variables $ [v(.); \vec a(.,.)]$  } 

\label{subsec_conjugc2.5diff}

For later purposes, it is important to stress here the consequences 
of the constitutive constraints on the empirical observables $[\rho(.);\vec j(.)]$ for their conjugated variables $ [v(.); \vec a(.,.)]$ :

(i) the normalization of Eq. \ref{rho1ptnormadiff} for the empirical density $\rho(.)$
means that only the inhomogeneities of the function $v(.)$ are really important.
If one adds an arbitrary constant $c$ to $v(.)$ everywhere
\begin{eqnarray}
 v_c(\vec x) \equiv v(\vec x)  +c
\label{vavecconstant}
\end{eqnarray}
then the corresponding contribution in the exponential of the generating function of Eq. \ref{genediffdef} 
will be only shifted by a constant term
\begin{eqnarray}
 \int d^d \vec x v_c(\vec x)  \rho(\vec x)=  \int d^d \vec x v(\vec x)  \rho(\vec x) +  c
\label{vavecconstantexp}
\end{eqnarray}

(ii) the constitutive stationarity constraints of Eq. \ref{divergencenulle} for the empirical current $\vec j (.)$
yields that if one adds the gradient of an arbitrary function $\phi(.)$ to $\vec a (\vec x) $
\begin{eqnarray}
 \vec a^{[\phi(.)]} (\vec x) \equiv  \vec a (\vec x) + \vec \nabla \phi(\vec x)
\label{abydifference}
\end{eqnarray}
then the corresponding contribution in the exponential of the generating function of Eq. \ref{genediffdef}
is unchanged using integration by parts to apply Eq. \ref{divergencenulle}
\begin{eqnarray}
\int d^d \vec x  \ \vec a^{[\phi(.)]} (\vec x). \vec j(\vec x)
&& = \int d^d \vec x  \ \vec a (\vec x). \vec j(\vec x)
+\sum_{\mu=1}^d  \int d^d \vec x \   \frac{ \partial  \phi(\vec x) }{\partial x_{\mu} }   j_{\mu} (\vec x)
\nonumber \\
&& =  \int d^d \vec x \  \vec a (\vec x). \vec j(\vec x)
-  \int d^d \vec x \ \phi(\vec x) \left[ \sum_{\mu=1}^d \frac{ \partial  j_{\mu} (\vec x)  }{\partial x_{\mu} }  \right]
=  \int d^d \vec x \  \vec a (\vec x). \vec j(\vec x)
   \label{abydifferenceexp}
\end{eqnarray}


\subsubsection{ Eigenvalue problem governing the generating function 
$Z_T^{[v(.);\vec a(.)]}(\vec x \vert \vec x_0)  $ for large $T$  } 

\label{subsec_eigenZdiff}

For large $T$, the leading behavior of the generating function of Eq. \ref{genediffdef}
\begin{eqnarray}
Z_T^{[v(.);\vec a(.)]}(\vec x \vert \vec x_0) \equiv 
  \opsimeq_{T \to + \infty} 
 e^{T G[v(.);\vec a(.)] } \ \ r^{[v(.);\vec a(.)]}(\vec x) \ \  l^{[v(.);\vec a(.)]}(\vec x_0)
\label{genediffdomin0}
\end{eqnarray}
involves the highest eigenvalue $G[v(.);\vec a(.)] $ of the following deformed Fokker-Planck generator
${\cal F}^{[v(.);\vec a(.)]} $ with respect to the initial generator of Eq. \ref{fokkerplanckgenerator}
\begin{eqnarray}
 {\cal F}^{[v(.);\vec a(.)]}  \equiv -v( \vec x)
  -   \sum_{\mu=1}^d \left( \frac{ \partial  }{\partial x_{\mu} } -  a_{\mu} ( \vec x)\right) 
  \left[  F_{\mu} (\vec x )  -D_{\mu} (\vec x)  \left( \frac{\partial }{ \partial x_{\mu}} -  a_{\mu} ( \vec x)  \right)
\right]
\label{fokkerplanckgeneratordeformed}
\end{eqnarray}
with the corresponding positive right 
eigenvector $ r^{[v(.);\vec a(.)]}(x)   \geq 0 $ 
\begin{eqnarray}
&& G[v(.);\vec a(.)]    r^{[v(.);\vec a(.)]}( \vec x)   = {\cal F}^{[v(.);\vec a(.)]}r^{[v(.);\vec a(.)]}( \vec x) 
\label{eigenrightdiff}
\\
&&  =   -v( \vec x) r^{[v(.);\vec a(.)]}( \vec x) 
   -  \sum_{\mu=1}^d \left( \frac{\partial }{\partial x_{\mu} }  -  a_{\mu} ( \vec x)  \right)  \left[ F_{\mu}[ \vec x ]  r^{[v(.);\vec a(.)]}( \vec x)  \right]
   +  \sum_{\mu=1}^d   \left( \frac{\partial }{\partial x_{\mu} }  -  a_{\mu} ( \vec x)  \right) \left[ D_{\mu}(\vec x)   \left( \frac{\partial }{\partial x_{\mu} }  -  a_{\mu} ( \vec x)  \right)r^{[v(.);\vec a(.)]}( \vec x) \right]
\nonumber 
\end{eqnarray}
and the positive left eigenvector $l^{[v(.);\vec a(.)]}(x)  \geq 0$ 
involving the adjoint deformed operator $\left({\cal F}^{[v(.);\vec a(.)]} \right)^{\dagger} $
\begin{eqnarray}
&&G[v(.);\vec a(.)]    l^{[v(.);\vec a(.)]}( \vec x)   
=\left({\cal F}^{[v(.);\vec a(.)]} \right)^{\dagger} l^{[v(.);\vec a(.)]}( \vec x) 
 \label{fokkerplanckAgeneratorkdaggereigen}
 \\
&&  =   -v( \vec x) l^{[v(.);\vec a(.)]}( \vec x) 
   +  \sum_{\mu=1}^d F_{\mu}[ \vec x ]   \left( \frac{\partial }{\partial x_{\mu} }  +  a_{\mu} ( \vec x)  \right) l^{[v(.);\vec a(.)]}( \vec x) 
   +   \sum_{\mu=1}^d   
 \left( \frac{\partial }{\partial x_{\mu} }  +  a_{\mu} ( \vec x)  \right) \left[ D_{\mu}(\vec x)   \left( \frac{\partial }{\partial x_{\mu} }  +  a_{\mu} ( \vec x)  \right)l^{[v(.);\vec a(.)]}( \vec x) \right] 
 \nonumber
\end{eqnarray}
satisfying the normalization
\begin{eqnarray}
1 = \langle l^{[v(.);\vec a(.)]}  \vert r^{[v(.);\vec a(.)]} \rangle 
= \int d^d x l^{[v(.);\vec a(.)]}(x)  r^{[v(.);\vec a(.)]}(x) 
 \label{Wdiffknorma}
\end{eqnarray}

The non-hermitian quantum Hamiltonian $H^{[v(.);\vec a(.)]} $ corresponding to the opposite of the deformed Fokker-Planck operator ${\cal F}^{[v(.);\vec a(.)]}  $ of Eq. \ref{fokkerplanckgeneratordeformed}
\begin{eqnarray}
H^{[v(.);\vec a(.)]} = - {\cal F}^{[v(.);\vec a(.)]} 
&&  = v( \vec x)
  +   \sum_{\mu=1}^d \left( \frac{ \partial  }{\partial x_{\mu} } -  a_{\mu} ( \vec x)\right) 
  \left[  F_{\mu} (\vec x )  -D_{\mu} (\vec x)  \left( \frac{\partial }{ \partial x_{\mu}} -  a_{\mu} ( \vec x)  \right)
\right]
\nonumber \\
&& = -  \sum_{\mu=1}^d 
 \bigg(  \frac{\partial }{ \partial x_{\mu}}  -  A^{[\vec a(.)]}_{\mu} ( \vec x ) \bigg) D_{\mu} (\vec x) \bigg(  \frac{\partial }{ \partial x_{\mu}}  -  A^{[\vec a(.)]}_{\mu} ( \vec x ) \bigg)
    + V^{[v(.)]}(\vec x)
\label{Hamiltoniandeformed}
\end{eqnarray}
then involves the deformed vector potential
\begin{eqnarray}
A^{[\vec a(.)]}_{\mu} ( \vec x ) \equiv A_{\mu} ( \vec x ) + a_{\mu} ( \vec x )= \frac{  F_{\mu} ( \vec x )}{2 D_{\mu} ( \vec x )}+a_{\mu} ( \vec x )
\label{vectorpotdeformed}
\end{eqnarray}
and the deformed scalar potential  
\begin{eqnarray}
V^{[v(.)]}(\vec x) \equiv V(\vec x) +v(x)   
 = \sum_{\mu=1}^d \left[ \frac{ F_{\mu}^2 (\vec x) }{ 4 D_{\mu} (\vec x)} +
 \frac{1}{2}  \frac{ \partial F_{\mu} (\vec x)}{\partial x_{\mu} } \right] +v(x)
\label{scalarpotdeformed}
\end{eqnarray}
with respect to the initial quantum Hamiltonian of Eq. \ref{Hamiltonian}.
So in this quantum mechanical language, the functions $[v(.);\vec a(.)]$ have a very simple interpretation,
since they deform additively the scalar and the vector potentials respectively.


\subsubsection{ Functional Legendre transform between the explicit rate function $ I_{2.5}  [ \rho(.),  \vec j  (.)]  $ 
and the eigenvalues $G[v(.);\vec a(.)]  $  } 

The explicit rate function $ I_{2.5}  [  \rho(.),  \vec j  (.)]  $ of Eq. \ref{rate2.25diff} governing the joint probability distribution
$p^{[2.5]}_T \left[  \rho(.),  \vec j  (.) \right] $ of Eq. \ref{ld2.5diff}
and the (usually non-explicit) eigenvalues $G[v(.);\vec a(.)]  $ governing the generating function $Z_T^{[v(.);\vec a(.)]}(\vec x \vert \vec x_0)  $ via Eq. \ref{genediffdomin0} are related via the following functional Legendre transformations \cite{chetrite_formal} :

(i) The generating function $Z_T^{[v(.);\vec a(.)]}(\vec x \vert \vec x_0)  $ of Eq. \ref{genediffdef}
can be computed from joint probability distribution
$p^{[2.5]}_T \left[  \rho(.),  \vec j  (.) \right] $ of Eq. \ref{ld2.5diff}
\begin{eqnarray}
&& Z_T^{[v(.);\vec a(.)]} = \int d\rho(.)  \int d \vec j(.) p^{[2.5]}_T[ \rho(.),  \vec j (.)] 
e^{ \displaystyle T \left[  -v(\vec x)  \rho(\vec x)   + \vec a (\vec x). \vec j(\vec x)
\right] }
\nonumber \\
&&  \opsimeq_{T \to + \infty} 
\int d\rho(.)  \int d \vec j(.) C_{[2.5]}[ \rho(.),  \vec j (.)] 
e^{ \displaystyle T \left[ 
  -v(\vec x)  \rho(\vec x)   + \vec a (\vec x). \vec j(\vec x)  -  I_{2.5}  [ \rho(.),  \vec j  (.)] \right] 
 }
 \opsimeq_{T \to + \infty}  
 e^{ \displaystyle T G[v(.);\vec a(.)] }
\label{geneddomin0saddlediff}
\end{eqnarray}
via the saddle-point method for large $T$ : so $G[G[v(.);\vec a(.)]] $
corresponds to the optimal value of the function in the exponential over the empirical density $\rho(.)$ and the empirical current $\vec j(.)$ satisfying the constitutive constraints $ C_{2.5} \left[ \rho(.),  \vec j (.) \right]$ of Eq. \ref{c2.25diff}
\begin{eqnarray}
G[v(.);\vec a(.)]
= \max_{\substack{\rho(.)  \text{ and }  \vec j(.)\\ \text{satisfying } C_{2.5} \left[ \rho(.),  \vec j (.)  \right]}}
 \left[ 
-v(\vec x)  \rho(\vec x)   + \vec a (\vec x). \vec j(\vec x)  -  I_{2.5}  [ \rho(.),  \vec j  (.)]\right] 
\label{legendrediff}
\end{eqnarray}
This optimization problem using the explicit expression of $ I_{2.5}  [  \rho(.),  \vec j  (.)]  $ of Eq. \ref{rate2.25diff}
and Lagrange multipliers to impose the constitutive constraints $ C_{2.5} \left[ \rho(.),  \vec j (.) \right]$ of Eq. \ref{c2.25diff}
leads to the eigenvalue problem for $G[v(.);\vec a(.)] $ described in the previous subsection \ref{subsec_eigenZdiff}.
Note that the eigenvalue $G[v(.);\vec a(.)] $ is usually not known explicitly 
in terms of the two arbitrary functions $[v(.);\vec a(.)]  $.

(ii) The reciprocal Legendre transformation of Eq. \ref{legendrediff} 
\begin{eqnarray}
I_{2.5}\left[ \rho(.),  \vec j  (.) \right]
= \max_{v(.)  \text{ and }  \vec a(.)}
 \left[ 
-v(\vec x)  \rho(\vec x)   + \vec a (\vec x). \vec j(\vec x)  -  G[v(.);\vec a(.)]
\right] 
\label{legendrerecidiff}
\end{eqnarray}
leads to the explicit expression of $ I_{2.5}  [  \rho(.),  \vec j  (.)]  $ of Eq. \ref{rate2.25diff}
using only that $G[v(.);\vec a(.)] $ is the eigenvalue for the eigenvalue problem described in the previous subsection \ref{subsec_eigenZdiff}
 (since  
the eigenvalues $G[v(.);\vec a(.)] $ are not known explicitly in terms of $[v(.);\vec a(.)]  $ in most applications).


\subsection{ Canonical conditionings with respect to the empirical density $\rho(.)$ and to the empirical current $\vec j(.)$ }

\subsubsection{ Conditioned Fokker-Planck generator ${\cal F}^{Cond[v(.);\vec a(.)]} $ associated to the deformed-generator ${\cal F}^{[v(.);\vec a(.)]} $} 

\label{subsec_directconddiff}

The conditioned Fokker-Planck generator ${\cal F}^{Cond[v(.);\vec a(.)]} $
associated to the deformed-generator ${\cal F}^{[v(.);\vec a(.)]} $ of Eq. \ref{fokkerplanckgeneratordeformed}
involves the same diffusion coefficients $D_{\mu}(\vec x)$ as the initial Fokker-Planck generator ${\cal F}$ of Eq. \ref{fokkerplanckgenerator}
\begin{eqnarray}
{\cal F}^{Cond[v(.);\vec a(.)]}
  =     -  \sum_{\mu=1}^d \frac{\partial }{\partial x_{\mu} }  \left[   F^{Cond[v(.);\vec a(.)]}_{\mu}(\vec x)  -  D_{\mu}(\vec x)   \frac{\partial }{\partial x_{\mu} }  \right]
\label{doobFPgenerator}
\end{eqnarray}
while the initial force $F_{\mu}(\vec x)$ is replaced by the conditioned force
\begin{eqnarray}
  F^{Cond[v(.);\vec a(.)]}_{\mu}(\vec x)
  = F_{\mu}(\vec x) +2 D_{\mu}(\vec x)  \bigg(    a_{\mu}(\vec x)      
      + \frac{ \partial   \ln \big[  l^{[v(.);\vec a(.)]} ( \vec x) \big] }{\partial x_{\mu}  } \bigg)
\label{forceDoobDiff}
\end{eqnarray}
that requires the knowledge of the left eigenvector $ l^{[v(.);\vec a(.)]} ( \vec x) $ of Eq. \ref{fokkerplanckAgeneratorkdaggereigen},
while the right eigenvector $r^{[v(.);\vec a(.)]} ( \vec x) $ of Eq. \ref{eigenrightdiff}
also appears in the 
corresponding conditioned steady state
\begin{eqnarray}
\rho^{Cond[v(.);\vec a(.)]}_*
  =   l^{[v(.);\vec a(.)]} ( \vec x)  r^{[v(.);\vec a(.)]} ( \vec x)
    \label{doobFPsteady}
\end{eqnarray}


\subsubsection{ Special cases $[v(.);\vec a(.)] $ with vanishing eigenvalue $G[v(.);\vec a(.)] =0$ and trivial left eigenvector $l^{[v(.);\vec a(.)]}( \vec x)=1 $}

For any given $\vec a(\vec x) $, one can choose the appropriate function $v^{[\vec a(.)]}(\vec x)$
corresponding to the vanishing eigenvalue $G[v(.);\vec a(.)] =0$ and 
to the trivial left eigenvector $l^{[v(.);\vec a(.)]}( \vec x)=1 $ in Eq. \ref{fokkerplanckAgeneratorkdaggereigen}
\begin{eqnarray}
   v^{[\vec a(.)]}(\vec x) =
     \sum_{\mu=1}^d F_{\mu}[ \vec x ]    a_{\mu} ( \vec x)  
    +  \sum_{\mu=1}^d   
 \left( \frac{\partial \left[ D_{\mu}(\vec x)   a_{\mu} ( \vec x)  \right]}{\partial x_{\mu} }  +  
 \left[ D_{\mu}(\vec x)   a_{\mu}^2 ( \vec x)  \right]  \right)  
  \label{alphaconservedasafunctionofbeta}
\end{eqnarray}
The corresponding conditioned force of Eq. \ref{forceDoobDiff}
reduces to
\begin{eqnarray}
  F^{Cond[v^{[\vec a(.)(.)]};\vec a(.)]}_{\mu}(\vec x)
  = F_{\mu}(\vec x) +2 D_{\mu}(\vec x)   a_{\mu}(\vec x)      
\label{forceDoobDiffconserved}
\end{eqnarray}

The corresponding deformed-generator ${\cal F}^{[v^{[\vec a(.)]}(.);\vec a(.)]} $ of Eq. \ref{eigenrightdiff}
\begin{eqnarray}
 {\cal F}^{[v^{[\vec a(.)]}(.);\vec a(.)]}
&& =  -v^{[\vec a(.)]}( \vec x) 
   -  \sum_{\mu=1}^d \left( \frac{\partial }{\partial x_{\mu} }  -  a_{\mu} ( \vec x)  \right)   F_{\mu}[ \vec x ]  
   +  \sum_{\mu=1}^d   \left( \frac{\partial }{\partial x_{\mu} }  -  a_{\mu} ( \vec x)  \right)  D_{\mu}(\vec x)   \left( \frac{\partial }{\partial x_{\mu} }  -  a_{\mu} ( \vec x)  \right)
\nonumber \\
&&  = 
   -  \sum_{\mu=1}^d  \frac{\partial }{\partial x_{\mu} }  \left[  F_{\mu}[ \vec x ]  +2 D_{\mu}(\vec x)   a_{\mu}(\vec x) 
   -   D_{\mu}(\vec x)  \frac{\partial }{\partial x_{\mu} }    
                      \right)
                 \nonumber \\
&&     \equiv  -  \sum_{\mu=1}^d  \frac{\partial }{\partial x_{\mu} }  \left[   F^{Cond[v^{[\vec a(.)(.)]};\vec a(.)]}_{\mu}(\vec x)
   -   D_{\mu}(\vec x)  \frac{\partial }{\partial x_{\mu} }    
                      \right)
                   =     {\cal F}^{Cond[v^{[\vec a(.)]}(.);\vec a(.)]}
\label{deformedcaseconserved}
\end{eqnarray}
then coincides with the conditioned Fokker-Planck generator ${\cal F}^{Cond[v^{[\vec a(.)]}(.);\vec a(.)]} $ 
of Eq. \ref{doobFPgenerator}
involving the conditioned force $F^{Cond[v^{[\vec a(.)(.)]};\vec a(.)]}_{\mu}(\vec x)
  = F_{\mu}(\vec x) +2 D_{\mu}(\vec x)   a_{\mu}(\vec x) $ of Eq. \ref{forceDoobDiffconserved}.
  
These special cases $[v^{[\vec a(.)]}(.);\vec a(.)]$ with
vanishing eigenvalues $G[v^{[\vec a(.)]}(.);\vec a(.)]  =0$ and 
 trivial left eigenvectors $l^{[v^{[\vec a(.)]}(.);\vec a(.)]}(\vec x)=1 $
 will be very useful to analyze the inverse problem discussed in the next Appendix.


\section{ Analysis of the inverse problem for diffusion processes  } 

\label{app_Inverse}

The goal of the present Appendix is to analyze the following inverse problem with respect to the direct canonical conditioning problem described in the subsection  \ref{subsec_directconddiff} of the previous Appendix \ref{app_Diffusion}.
The ideas are the same as in the section \ref{sec_Inverse} of the main text concerning Markov jump processes,
but there are some important technical differences that need to be emphasized.

Here one wishes to determine whether a given Fokker-Planck generator ${\mathring {\cal F}}$
can be interpreted as the canonical conditioned generator ${\cal F}^{Cond[v(.); \vec a(.)]} $ of 
Eq. \ref{doobFPgenerator}
of the given initial Fokker-Planck generator ${\cal F}$ with appropriate functions $ [v(.); \vec a(.)]$
\begin{eqnarray}
{\mathring {\cal F}} = {\cal F}^{Cond[v(.);\vec a(.)]}
  =     -  \sum_{\mu=1}^d \frac{\partial }{\partial x_{\mu} }  \left[   F^{Cond[v(.);\vec a(.)]}_{\mu}(\vec x)  -  D_{\mu}(\vec x)   \frac{\partial }{\partial x_{\mu} }  \right]
\label{doobFPgeneratorinverse}
\end{eqnarray}

An obvious condition is that the two Fokker-Planck generators  ${\mathring {\cal F}} $ and ${\cal F} $ should have the same diffusion coefficients 
$D_{\mu}(\vec x)$.
Then the force ${\mathring F}_{\mu} (\vec x) $ should coincide with the conditional force $F^{Cond[v(.);\vec a(.)]}_{\mu}(\vec x) $ of Eq. \ref{forceDoobDiff}
\begin{eqnarray}
{\mathring F}_{\mu} (\vec x)=  F^{Cond[v(.);\vec a(.)]}_{\mu}(\vec x)
  = F_{\mu}(\vec x) +2 D_{\mu}(\vec x)  \bigg(    a_{\mu}(\vec x)      
      + \frac{ \partial   \ln \big[  l^{[v(.);\vec a(.)]} ( \vec x) \big] }{\partial x_{\mu}  } \bigg)
\label{forceDoobDiffinverse}
\end{eqnarray}
The goal is then to determine the functions $ [v(.); \vec a(.)]$
satisfying Eq. \ref{forceDoobDiffinverse} in terms of the two
given forces ${\mathring F}_{\mu} (\vec x) $ and $F_{\mu}(\vec x)  $.


\subsection{ Simple solution $[{\mathring v}(.); {\mathring a}_{\mu}(.)]$ 
with vanishing eigenvalue $G [{\mathring v}(.); {\mathring a}_{\mu}(.)] =0$ 
and left eigenvector $l^{[{\mathring v}(.); {\mathring a}_{\mu}(.)]} ( \vec x) =1 $ } 

If one chooses ${\mathring v}(\vec x) =v^{[ {\mathring a}_{\mu}(.)]}(\vec x)$ of Eq. \ref{alphaconservedasafunctionofbeta},
then the functions $  {\mathring a}_{\mu} $ obtained from Eq. \ref{forceDoobDiffconserved}
\begin{eqnarray}
{\mathring a}_{\mu}(\vec x)     =\frac{  {\mathring F}_{\mu}(\vec x) - F_{\mu}(\vec x) }{2 D_{\mu}(\vec x)  }
\equiv  {\mathring A}_{\mu}(\vec x) - A_{\mu}(\vec x)
\label{vectorpotsoluconserved}
\end{eqnarray}
corresponds to the difference between the two vector potentials ${\mathring A}_{\mu}(\vec x) $ and $A_{\mu}(\vec x) $
associated to the two forces ${\mathring F}_{\mu}(\vec x) $ and $F_{\mu}(\vec x) $ via Eq. \ref{vectorpot}.

Plugging Eq. \ref{vectorpotsoluconserved} into Eq. \ref{alphaconservedasafunctionofbeta}
then yields 
\begin{eqnarray}
 {\mathring v}(\vec x)   
=v^{[ {\mathring a}_{\mu}(.)]}(\vec x)
&& =     \sum_{\mu=1}^d F_{\mu}[ \vec x ]    {\mathring a}_{\mu} ( \vec x)  
   +   \sum_{\mu=1}^d   
 \left( \frac{\partial }{\partial x_{\mu} }  +  {\mathring a}_{\mu} ( \vec x)  \right) \left[ D_{\mu}(\vec x)   {\mathring a}_{\mu} ( \vec x)  \right] 
\nonumber \\
   && =
     \sum_{\mu=1}^d F_{\mu}[ \vec x ]   \frac{  {\mathring F}_{\mu}(\vec x) - F_{\mu}(\vec x) }{2 D_{\mu}(\vec x)  }
   +   \sum_{\mu=1}^d   
 \left( \frac{\partial }{\partial x_{\mu} }  + \frac{  {\mathring F}_{\mu}(\vec x) - F_{\mu}(\vec x) }{2 D_{\mu}(\vec x)  } \right) \left[  \frac{  {\mathring F}_{\mu}(\vec x) - F_{\mu}(\vec x) }{2  } \right] 
 \nonumber \\
 && =  
      \sum_{\mu=1}^d   
 \frac{\partial }{\partial x_{\mu} }  \left[  \frac{  {\mathring F}_{\mu}(\vec x) - F_{\mu}(\vec x) }{2  } \right] 
 +   \sum_{\mu=1}^d   
 \frac{  {\mathring F}_{\mu}^2(\vec x) - F^2_{\mu}(\vec x) }{4 D_{\mu}(\vec x)  } 
  \nonumber \\
 && \equiv  {\mathring V}(\vec x) - V(\vec x)
  \label{scalarpotsoluconserved}
\end{eqnarray}
that ${\mathring v}(\vec x)   $ corresponds to the difference between the two scalar potentials ${\mathring V}(\vec x) $ and $V(\vec x) $
associated to the two forces ${\mathring F}_{\mu}(\vec x) $ and $F_{\mu}(\vec x) $ via Eq. \ref{scalarpot}


\subsection{ Other solutions $ [v(.); \vec a(.)]$ corresponding to other eigenvalues $G$ and other eigenvectors $l(.)$ } 

The other solutions $ [v(.); \vec a(.)]$ for the inverse problem of Eq. \ref{forceDoobDiffinverse}
can be discussed using the simple solution of Eqs \ref{vectorpotsoluconserved}
and \ref{scalarpotsoluconserved} as follows.
Plugging Eq. \ref{vectorpotsoluconserved}
into Eq. \ref{forceDoobDiffinverse}
yields that $ a_{\mu}(\vec x)    $ can differ from the simple solution $ {\mathring a}_{\mu} (\vec x) $
only via the following derivative of the logarithm of the left eigenvector $ l^{[v(.);\vec a(.)]} ( \vec x) $
\begin{eqnarray}
  a_{\mu}(\vec x)       = {\mathring a}_{\mu} (\vec x)
  - \frac{ \partial   \ln \big[  l^{[v(.);\vec a(.)]} ( \vec x) \big] }{\partial x_{\mu}  } 
\label{betaDiffinversegene}
\end{eqnarray}

Plugging this into
the eigenvalue Eq. \ref{fokkerplanckAgeneratorkdaggereigen} then yields
\begin{eqnarray}
G[v(.);\vec a(.)]  
&&  =   -v( \vec x) 
   +  \sum_{\mu=1}^d F_{\mu}[ \vec x ]   {\mathring a}_{\mu} (\vec x) 
   +   \sum_{\mu=1}^d   
 \left( \frac{\partial \left[ D_{\mu}(\vec x)  {\mathring a}_{\mu} (\vec x)
 \right]}{\partial x_{\mu} }  +  a_{\mu} ( \vec x) D_{\mu}(\vec x)  {\mathring a}_{\mu} (\vec x) \right)  
   +   \sum_{\mu=1}^d   D_{\mu}(\vec x)  {\mathring a}_{\mu} (\vec x)
\left[ {\mathring a}_{\mu} (\vec x) -   a_{\mu}(\vec x)  \right]
\nonumber \\
&& =  -v( \vec x) 
   +  \sum_{\mu=1}^d F_{\mu}[ \vec x ]   {\mathring a}_{\mu} (\vec x) 
   +   \sum_{\mu=1}^d   
 \frac{\partial \left[ D_{\mu}(\vec x)  {\mathring a}_{\mu} (\vec x)
 \right]}{\partial x_{\mu} } 
   +   \sum_{\mu=1}^d   D_{\mu}(\vec x)  {\mathring a}^2_{\mu} (\vec x)
   \nonumber \\
&&  =-v( \vec x) + {\mathring v}(\vec x) 
 \label{eigenleftforring}
\end{eqnarray}
where we have used the first line of Eq. \ref{scalarpotsoluconserved}
in order to recognize the simple solution ${\mathring v}(\vec x)  $.
So the difference between $  v( \vec x)$ and ${\mathring v}(\vec x) $ 
 reduces to the constant given by the eigenvalue $G[v(.);\vec a(.)]    $
\begin{eqnarray}
 v( \vec x) && = {\mathring v}(\vec x) - G [v(.);\vec a(.)]
 \label{alphagendiff}
\end{eqnarray}
Note that the relations of Eqs \ref{betaDiffinversegene} and \ref{alphagendiff} between the different solutions
can be understood from the properties of $ [v(.); \vec a(.)]$ discussed in subsection \ref{subsec_conjugc2.5diff}.

In summary, the choice of the eigenvalue $G$ determines the function
 $v( \vec x)$ via Eq. \ref{alphagendiff}
\begin{eqnarray}
v^{[G]}(\vec x) && = -G + {\mathring v}(\vec x)= - G   + {\mathring V}(\vec x) - V(\vec x)
\nonumber \\
&& =  - G +  \sum_{\mu=1}^d   
 \frac{\partial }{\partial x_{\mu} }  \left[  \frac{  {\mathring F}_{\mu}(\vec x) - F_{\mu}(\vec x) }{2  } \right] 
 +   \sum_{\mu=1}^d   
 \frac{  {\mathring F}_{\mu}^2(\vec x) - F^2_{\mu}(\vec x) }{4 D_{\mu}(\vec x)  } 
\label{alphaGsolugene}
\end{eqnarray}
while the choice of the positive left eigenvector $l(.)>0$ determines the function 
$ a_{\mu}(\vec x)    $ via Eq. \ref{betaDiffinversegene}
\begin{eqnarray}
  a_{\mu}^{[l(.)]}(\vec x)       = {\mathring a}_{\mu} (\vec x)
  - \frac{ \partial   \ln \big[  l ( \vec x) \big] }{\partial x_{\mu}  } 
  = \frac{  {\mathring F}_{\mu}(\vec x) - F_{\mu}(\vec x) }{2 D_{\mu}(\vec x)  } 
    - \frac{ \partial   \ln \big[  l ( \vec x) \big] }{\partial x_{\mu}  } 
\label{betaDiffinversegenel}
\end{eqnarray}

In the two remaining Appendices, this general solution for the inverse problem
is illustrated with examples on the one-dimensional ring (Appendix \ref{app_Ring})
and in the full space $R^d$ in dimension $d$ (Appendix \ref{app_Diffd}).


\section{ Application to Fokker-Planck dynamics on the one-dimensional periodic ring  }

\label{app_Ring}

In this Appendix, we focus on the Fokker-Planck generator on the one-dimensional periodic ring of length $L$
\begin{eqnarray}
{\mathring {\cal F} } \equiv  -    \frac{ \partial  }{\partial x } \left[  {\mathring F} ( x )  -D (x)  \frac{\partial }{ \partial x}   
\right]
\label{fokkerplanckgeneratord1}
\end{eqnarray}
 involving the arbitrary force ${\mathring F}(x)$ 
and the arbitrary diffusion coefficient $D(x)$,
whose large deviations properties at various levels have been discussed in \cite{c_lyapunov}.

The simplest Fokker-Planck generator with the same diffusion coefficient $D(x)$ is 
the model without force $F(x)=0$ of generator
\begin{eqnarray}
 {\cal F}  \equiv  -    \frac{ \partial  }{\partial x } D (x)  \frac{\partial }{ \partial x}   
\label{fokkerplanckgeneratord1noforce}
\end{eqnarray}
that satisfies detailed-balance with respect to its uniform steady state $P(x)=\frac{1}{L}$


\subsection{ Direct problem : canonical conditioning of the diffusion with arbitrary $D(x)$ and zero force $F(x)=0$ } 

On the periodic ring of length $L$, the vanishing-divergence constraint of Eq. \ref{divergencenulle}
for the empirical current $j(x)$ yields that it is uniform along the ring
\begin{eqnarray}
j(x)=j  
\label{jringconserved}
\end{eqnarray}
As a consequence, the conditioning observable of Eq. \ref{genediffdef} parametrized by the two functions $[v(x);a(x)]$
\begin{eqnarray}
\int_0^L dx \left(    a (x)  j(x) -v(x)  \rho(x)   + \right) = j \Phi - \int_0^L dx v(x)  \rho(x)   \ \ \text{ with } \ \ 
\Phi \equiv \int_0^L dx a (x)
\label{jringconservedobs}
\end{eqnarray}
involves the empirical density $\rho(x)$ and the empirical current $j$, whose conjugated parameter is the 
total flux $\Phi$ given by the integration of the vector potential $a(x)$ along the ring.
One recognizes the famous Aharonov-Bohm problem,
where only the total magnetic flux through the ring is relevant (see \cite{c_lyapunov} for more details in the present context of large deviations),

One needs to solve the eigenvalue Eq. \ref{fokkerplanckAgeneratorkdaggereigen} for the positive periodic left eigenvector $ l^{[v(.); a(.)]}(  x) >0$
\begin{eqnarray}
&&G[v(.); a(.)]    l^{[v(.); a(.)]}(  x)   
  =   -v(  x) l^{[v(.); a(.)]}(  x) 
   +
 \left( \frac{\partial }{\partial x }  +  a (  x)  \right) \left[ D( x)   \left( \frac{\partial }{\partial x }  +  a (  x)  \right)l^{[v(.); a(.)]}(  x) \right] 
\label{eigenleftdiffring}
\end{eqnarray}
in order to compute the conditioned force via Eq. \ref{forceDoobDiff}
\begin{eqnarray}
  F^{Cond[v(.);a(.)]}(x)
  = 2 D(x)  \bigg(    a(x)            + \frac{ \partial   \ln \big[  l^{[v(.);a(.)]} ( x) \big] }{\partial x  } \bigg)
\label{forceDoobDiffring}
\end{eqnarray}

While the vector potential $a(x)$ on the ring can always be taken into account via boundary conditions
after a gauge transformation,
it is not possible to solve explicitly the eigenvalue Eq. \ref{eigenleftdiffring}
for arbitrary scalar potentials $v(x)$, even when the diffusion coefficient $D(x)$ is constant.


\subsection{ Inverse problem : functions $[v(x),a(x)]$ to produce 
the given force ${\mathring F}(x)=F^{Cond[v(.);a(.)]}(x)$} 

The general solution described in Appendix \ref{app_Inverse}
can be applied to the two Fokker-Planck generators of Eqs \ref{fokkerplanckgeneratord1}
and \ref{fokkerplanckgeneratord1noforce} as follows:

(i) the simple solution of Eqs \ref{vectorpotsoluconserved} and \ref{scalarpotsoluconserved} reduces to
\begin{eqnarray}
{\mathring a}(x)  &&   =\frac{  {\mathring F}(x)  }{2 D(x)  }
\nonumber \\
{\mathring v}(x)   && =  
       \frac{  {\mathring F}'(x)  }{2  } 
 +    \frac{  {\mathring F}^2(x)  }{4 D(x)  } 
  \label{alphaconservedasafunctionofbetaRingdiff}
\end{eqnarray}

(ii) the other solutions can be parametrized by the 
eigenvalue $G$ that determines the scalar potential via Eq. \ref{alphaGsolugene}
\begin{eqnarray}
v^{[G]}( x) && = -G + {\mathring v}(x) =  - G +    \frac{  {\mathring F}'(x)  }{2  } 
 +    \frac{  {\mathring F}^2(x)  }{4 D(x)  } 
\label{alphaGsolugenering}
\end{eqnarray}
and by the positive periodic left eigenvector $l(.)>0$ that determines the vector potential 
 via Eq. \ref{betaDiffinversegenel}
\begin{eqnarray}
  a^{[l(.)]}( x)       = {\mathring a} ( x)
  - \frac{ \partial   \ln \big[  l (  x) \big] }{\partial x  } 
  =\frac{  {\mathring F}(x)  }{2 D(x)  }
 - \frac{ \partial   \ln \big[  l (  x) \big] }{\partial x  } 
\label{betaDiffinversegenering}
\end{eqnarray}
So the integration of this vector potential along the ring
that corresponds to the flux $\Phi$ conjugated to the empirical current $j(x)=j$ in Eq. \ref{jringconservedobs}
\begin{eqnarray}
\Phi = \int_0^L dx a^{[l(.)]}( x)      = \int_0^L dx \frac{  {\mathring F}(x)  }{2 D(x)  } 
\label{betaDiffinversegeneringflux}
\end{eqnarray}
is fixed once the force ${\mathring F}(x) $ is given.
One can thus distinguish two cases:

(a) When Eq. \ref{betaDiffinversegeneringflux} vanishes $ \int_0^L dx \frac{  {\mathring F}(x)  }{2 D(x)  }=0 $,
then the Fokker-Planck generator of{fokkerplanckgeneratord1} corresponds to an equilibrium dynamics.
One can choose the left eigenvector $l(.)$ to make the vector potential identically vanishing $ a^{[l(.)]}( x)  =0 $, so that the conditioning will be only with respect to the empirical density $\rho(.)$ 
via the conjugate parameter $ v^{[G]}( x)$ of Eq. \ref{alphaGsolugenering}.

(b) When Eq. \ref{betaDiffinversegeneringflux} does not vanish, $ \int_0^L dx \frac{  {\mathring F}(x)  }{2 D(x)  } \ne 0 $, then the Fokker-Planck generator of{fokkerplanckgeneratord1} corresponds to an out-of-equilibrium dynamics. The conditioning then involves both the empirical current $j(x)=j$ along the ring
via the conjugate parameter of Eq. \ref{betaDiffinversegeneringflux}, and the empirical density $\rho(.)$ 
via the conjugate parameter $ v^{[G]}( x)$ of Eq. \ref{alphaGsolugenering}.


\section{ Application to non-equilibrium Fokker-Planck dynamics in the full d-dimensional space $R^d$ }

\label{app_Diffd}

In this Appendix, we focus on the non-equilibrium Fokker-Planck dynamics in the full d-dimensional space $R^d$
converging towards some normalizable steady state,
in order to obtain its canonical conditioning interpretation with respect to the detailed-balance Fokker-Planck dynamics
that converges towards the same steady state.

\subsection{ Decomposition of the force $ {\mathring F}_{\mu}(\vec x) = {\mathring F}^{rev}_{\mu}(\vec x)+ {\mathring F}^{irr}_{\mu}(\vec x) $ into its reversible and irreversible contributions }

When the Fokker-Planck dynamics involving the force 
$ {\mathring F}_{\mu}(\vec x)$ and the diffusion coefficients $D_{\mu}(\vec x)$
converges towards the normalizable steady state ${\mathring \rho}^*(\vec x) $
associated to a non-vanishing steady current ${\mathring J}^*_{\mu}(\vec x) $
\begin{eqnarray}
 {\mathring J}^*_{\mu}(\vec x) \equiv  {\mathring \rho}^*(\vec x ) {\mathring F}_{\mu}(\vec x)
  - D_{\mu}(\vec x)  \frac{ \partial {\mathring \rho}^*(\vec x )}{\partial x_{\mu}}
\label{jsteadyring}
\end{eqnarray}
 it is standard to decompose the force ${\mathring F}_{\mu}(\vec x) $ into its reversible and irreversible contributions
\begin{eqnarray}
  {\mathring F}_{\mu}(\vec x) = {\mathring F}^{rev}_{\mu}(\vec x)+ {\mathring F}^{irr}_{\mu}(\vec x)
\label{revirrev}
\end{eqnarray} 
The reversible contribution ${\mathring F}^{rev}_{\mu}(\vec x) $ 
is the force that would produce a vanishing steady current in Eq. \ref{jsteadyring}
\begin{eqnarray}
{\mathring F}^{rev}_{\mu}(\vec x) 
= D_{\mu}(\vec x)  \frac{ \partial \ln \big[{\mathring \rho}^*(\vec x ) \big]}{\partial x_{\mu}}
\label{forceRev}
\end{eqnarray}
so that the remaining irreversible contribution ${\mathring F}^{irr}_{\mu}(\vec x)
= {\mathring F}_{\mu}(\vec x) - {\mathring F}^{rev}_{\mu}(\vec x) $
is directly responsible for the non-vanishing steady current of Eq. \ref{jsteadyring} via
\begin{eqnarray}
 {\mathring J}^*_{\mu}(\vec x)  =  {\mathring \rho}^*(\vec x ) {\mathring F}^{irr}_{\mu}(\vec x)
 \label{jsteadyirrev}
\end{eqnarray}
The constraint of vanishing divergence for this steady current $ {\mathring J}^*_{\mu}(\vec x) $
\begin{eqnarray}
0 =   \sum_{\mu=1}^d  \frac{ \partial {\mathring J}^*_{\mu} (\vec x ) ]}{\partial x_{\mu}}
=  \sum_{\mu=1}^d  \frac{ \partial [ {\mathring \rho}^*(\vec x ) {\mathring F}^{irr}_{\mu}(\vec x) ]}{\partial x_{\mu}}
=\sum_{\mu=1}^d  \frac{ \partial  {\mathring \rho}^*(\vec x )   }{\partial x_{\mu}} . {\mathring F}^{irr}_{\mu}(\vec x)
+ {\mathring \rho}^*(\vec x ) \sum_{\mu=1}^d  \frac{ \partial   {\mathring F}^{irr}_{\mu}(\vec x)  }{\partial x_{\mu}}
 \label{divjsteadyirrev}
\end{eqnarray}
translates into the following constraint for the irreversible force using Eq. \ref{forceRev}
\begin{eqnarray}
 \sum_{\mu=1}^d  \frac{ \partial [  {\mathring F}^{irr}_{\mu}(\vec x)  ]}{\partial x_{\mu}}
  = -  \sum_{\mu=1}^d  \frac{ \partial \ln [ {\mathring \rho}^*(\vec x )   ]}{\partial x_{\mu}}  {\mathring F}^{irr}_{\mu}(\vec x)
  = -  \sum_{\mu=1}^d  \frac{ {\mathring F}^{rev}_{\mu}(\vec x)  {\mathring F}^{irr}_{\mu}(\vec x) }{ D_{\mu}(\vec x)} 
 \label{divforceirrev}
\end{eqnarray}


\subsection{ Inverse problem : $ {\mathring F}_{\mu}(\vec x) = {\mathring F}^{rev}_{\mu}(\vec x)+ {\mathring F}^{irr}_{\mu}(\vec x) $ as conditioning of the dynamics associated to $F(\vec x)={\mathring F}^{rev}_{\mu}(\vec x) $ }

Let us analyze how the non-equilibrium Fokker-Planck dynamics with force 
$ {\mathring F}_{\mu}(\vec x) = {\mathring F}^{rev}_{\mu}(\vec x)+ {\mathring F}^{irr}_{\mu}(\vec x) $
described above can be considered as the canonical conditioning of the equilibrium dynamics
involving only the reversible force  
\begin{eqnarray}
F(\vec x)={\mathring F}^{rev}_{\mu}(\vec x)
\label{revasreference}
\end{eqnarray}
with appropriate functions $[v(.);\vec a(.)]$. 

The general solution of Appendix \ref{app_Inverse}
can be applied to the present example as follows :

(i) the simple solution of Eqs \ref{vectorpotsoluconserved} and \ref{scalarpotsoluconserved} 
yields that the vector potential is directly related to the irreversible force ${\mathring F}^{irr}_{\mu}(\vec x) $
\begin{eqnarray}
{\mathring a}_{\mu}(\vec x)  &&   =\frac{  {\mathring F}_{\mu}(\vec x) - F_{\mu}(\vec x) }{2 D_{\mu}(\vec x)  }
=\frac{  {\mathring F}^{irr}_{\mu}(\vec x)   }{2 D_{\mu}(\vec x)  }
  \label{soluirrsimplea}
\end{eqnarray}
while the vector potential can be also rewritten in terms of  the irreversible force ${\mathring F}^{irr}_{\mu}(\vec x) $
 using Eq. \ref{divforceirrev}
\begin{eqnarray}
 {\mathring v}(\vec x)   
&& =     \sum_{\mu=1}^d   
 \frac{\partial }{\partial x_{\mu} }  \left[  \frac{   {\mathring F}^{irr}_{\mu}(\vec x)  }{2  } \right] 
 +   \sum_{\mu=1}^d   
 \frac{  \left[ F_{\mu}(\vec x)+ {\mathring F}^{irr}_{\mu}(\vec x) \right]^2 - F^2_{\mu}(\vec x) }{4 D_{\mu}(\vec x)  } 
\nonumber \\
&& =  \sum_{\mu=1}^d   
 \frac{\partial }{\partial x_{\mu} }  \left[  \frac{   {\mathring F}^{irr}_{\mu}(\vec x)  }{2  } \right] 
 +   \sum_{\mu=1}^d   
 \frac{   2 F_{\mu}(\vec x){\mathring F}^{irr}_{\mu}(\vec x)+ [{\mathring F}^{irr}_{\mu}(\vec x)]^2  }{4 D_{\mu}(\vec x)  }
\nonumber \\
&&  = \sum_{\mu=1}^d    \frac{    [{\mathring F}^{irr}_{\mu}(\vec x)]^2  }{4 D_{\mu}(\vec x)  }
  \label{soluirrsimplev}
\end{eqnarray}

(ii) the other solutions can be parametrized by the 
eigenvalue $G$ that determines the scalar potential via Eq. \ref{alphaGsolugene}
\begin{eqnarray}
v^{[G]}(\vec x) && = -G + {\mathring v}(\vec x)= - G  
+ \sum_{\mu=1}^d    \frac{    [{\mathring F}^{irr}_{\mu}(\vec x)]^2  }{4 D_{\mu}(\vec x)  }
\label{alphaGsolugeneirr}
\end{eqnarray}
and by the positive left eigenvector $l(.)>0$ that determines the vector potential 
$ \vec a( x)    $ via Eq. \ref{betaDiffinversegenel}
\begin{eqnarray}
  a_{\mu}^{[l(.)]}(\vec x)       = {\mathring a}_{\mu} (\vec x)
  - \frac{ \partial   \ln \big[  l ( \vec x) \big] }{\partial x_{\mu}  } 
  = \frac{  {\mathring F}^{irr}_{\mu}(\vec x)   }{2 D_{\mu}(\vec x)  }
   - \frac{ \partial   \ln \big[  l ( \vec x) \big] }{\partial x_{\mu}  } 
  \label{betaDiffinversegenelirr}
\end{eqnarray}


\end{document}